\documentclass[%
 reprint,
superscriptaddress,
 amsmath,amssymb,
 aps,
]{revtex4-2}

\usepackage{graphicx}
\usepackage{dcolumn}
\usepackage{bm}
\usepackage[version=3]{mhchem} 

\begin{document}

\preprint{APS/123-QED}


\title{Metal halide thermoelectrics: prediction of high-performance \ce{CsCu2I3}}

\author{Jong Woong Park}
\affiliation{
Department of Materials Science and Engineering, Yonsei University, Seoul 03722, South Korea}%
\author{Young-Kwang Jung}%
 \email{yj359@cam.ac.uk; a.walsh@imperial.ac.uk}
 \affiliation{
Department of Chemical Engineering and Biotechnology, University of Cambridge, Cambridge CB3 0AS, UK}%
\author{Aron Walsh}
\email{yj359@cam.ac.uk; a.walsh@imperial.ac.uk}
\affiliation{Department of Materials, Imperial College London, Exhibition Road, London SW7 2AZ, UK}%
\affiliation{Department of Physics, Ewha Womans University, Seoul 03760, South Korea}%

\date{\today}

\begin{abstract}
Thermoelectric devices can directly convert waste heat into electricity, which makes them an important clean energy technology. The underlying materials performance can be evaluated by the dimensionless figure of merit $ZT$. Metal halides are attractive candidates due to their chemical flexibility and ease of processing; however, the maximum $ZT$ realized ($ZT$ = 0.15) falls far below the level needed for commercialization ($ZT$ $>$ 1). Using a first-principles procedure we assess the thermoelectric potential of copper halide \ce{CsCu2I3}, which features 1D Cu--I connectivity. The n-type crystal is predicted to exhibit a maximum $ZT$ of $2.2$ at $600$\,K along the $b$-axis. The strong phonon anharmonicity of this system is shown by locally stable non-centrosymmetric $Amm2$ structures that are averaged to form the observed centrosymmetric $Cmcm$ space group. Our work provides insights into the structure--property relations in metal halide thermoelectrics and suggests a path forward to engineer higher-performance heat-to-electricity conversion.
\end{abstract}

\maketitle


\section{\label{sec:level1}Introduction}

%
Thermoelectric materials -- which enable the direct conversion of waste heat to electricity -- have received great attention as one of the most promising renewable energy technologies \cite{Bell2008}.
The performance of thermoelectric materials is evaluated by the dimensionless figure of merit, $ZT$:
\begin{equation}
  ZT=\frac{S^2 \sigma T}{\kappa_{\rm elec} + \kappa_{\rm latt}}
\label{eqn_ZT}
\end{equation}
where $S$ is the Seebeck coefficient, $\sigma$ is the electrical conductivity, $T$ is the temperature, $\kappa_{\rm elec}$ is the electronic thermal conductivity, and $\kappa_{\rm latt}$ is the lattice thermal conductivity (power factor, $PF$, is defined as $S^2\sigma$).
Due to the recent progress in calculation methods for electron (hole) and phonon transport in solid crystals, computational studies have been made to discover novel compounds that possess a high $ZT$ \cite{Jain2016, Gorai2017, Spooner2021, Deng2021}.
Although the trade-off effect between the parameters that control $ZT$ makes the optimization challenging, search for an intrinsically low $\kappa_{\rm latt}$ material is still crucial to maximize performance\cite{Hao2019}.
In other words, high thermoelectric performance requires the phonons to be disrupted like in a glass but the electrons to have a high mobility like in crystalline semiconductors (i.e. phonon-glass electron-crystal) \cite{Rowe2010}.

\subsection{\label{sec:level2}Metal halide thermoelectrics}

Metal halides have been studied for their uses in various applications including solar cells \cite{Kojima2009, Lee2014}, light-emitting diodes \cite{Tan2014, Jun2018}, and memristors \cite{Xiao2015, Yang2021}.
These materials are also known to have an intrinsically `ultra-low' $\kappa_{\rm latt}$ ($< 1$\,W/m$\cdot$K) \cite{Pisoni2014, Lee2017, Haque2019}, and several reports have started to suggest their potential for thermoelectric applications (Table\,\ref{table_1}).

%
\begin{table*}
\caption{\label{table_1}Representative works on metal halides studies for thermoelectric devices with their $\kappa_{\rm latt}$ and $ZT$ values. The mentioned values are for room temperature unless indicated otherwise.}
\begin{ruledtabular}
\begin{tabular}{cccccc}
                     & Compound                                                        & $\kappa_{\rm latt}$ (W/m$\cdot$K) & $ZT$                  & $\kappa$ measurement & Ref.                                 \\ \hline
  & \ce{CsSnI3}                                    & 0.38                              & 0.11 (320 K)          & NW thermometry       & \cite{Lee2017}      \\
                     & \ce{CsSnI3}                                    & 0.28                              & 0.08 (323 K)          & 3-$\omega$ method    & \cite{Tang2022}     \\
                     & \ce{CsSnI3}                                    & 0.60                              & 0.15 (550 K)          & Laser flash method   & \cite{Xie2020}      \\
                     & \ce{CsSnI3} + \ce{SnCl2} 1\%  & 0.38                              & 0.14 (345 K)          & 3-$\omega$ method    & \cite{Liu2019}      \\
                     & \ce{CsSnI3} + \ce{PbI2} 0.5\% &                                   & 0.14 (523 K)          & Laser flash method   & \cite{Yu2022}       \\
                     & CsSn$_{0.8}$Ge$_{0.2}$I$_3$                                     &                                   & 0.12 (473 K)          & Laser flash method   & \cite{Qian2020}     \\
                    Conventional perovskites & \ce{CsSnBr3}                                   & 0.64                              &                       & Laser flash method   & \cite{Xie2020}      \\
                     & \ce{CsPbI3}                                    & 0.45                              &                       & NW thermometry       & \cite{Lee2017}      \\
                     & \ce{CsPbBr3}                                   & 0.42                              &                       & NW thermometry       & \cite{Lee2017}      \\
                     & \ce{CsPbBr3}                                   & 0.44 (tot.)                       &                       & 3-$\omega$ method    & \cite{Haeger2019}   \\
                     & \ce{CsPbBr3}                                   & 0.46 (tot.)                       &                       & FDTR                 & \cite{Elbaz2017}    \\
                     & \ce{CsPbCl3}                                   & 0.49 (tot.)                       &                       & 3-$\omega$ method    & \cite{Haeger2020}   \\
        & \ce{Cs2SnI6}                                   & 0.29                              &                       & Laser flash method   & \cite{Bhui2022}     \\
\hline
  & \ce{Cs2SnI2Cl2}                                & 0.60 (300 K)                      & \multicolumn{1}{l}{}  & Laser flash method   & \cite{Zeng2022}                  \\
                     & \ce{CsPb2Br5}                                  & 0.32 (tot.)                       &                       & 3-$\omega$ method    & \cite{Haeger2019}   \\
                     & \ce{Cs2PbI2Cl2}                                & 0.45 (300 K)                      &                       & Laser flash method   & \cite{Zeng2022}                                     \\
Low-D metal halides & \ce{Cs2PbI2Cl2}                                & 0.37 (295 K)                      &                       & Laser flash method   & \cite{Acharyya2020} \\
                     & \ce{Cs4PbCl6}                                  & 0.30 (tot.)                       &                       & 3-$\omega$ method    & \cite{Haeger2020}   \\
                     & \ce{Cs3Cu2I5}                                  & 0.02                              & 2.6 (600 K)           & Phono3py (theory)    & \cite{Jung2021}     \\
                     & \ce{CsCu2I3}                                   & 0.05 (300 K)                      & 0.4 (300 K)           & Phono3py (theory)    & This study                           \\
                     &                                                                 & 0.02 (600 K, $b$-axis)              & 2.2 (600 K, $b$-axis) &                      &                                     
\end{tabular}
\end{ruledtabular}
\end{table*}

However, the highest maximum $ZT$ of $0.15$ achieved from a halide perovskite \ce{CsSnI3} \cite{Xie2020} is far to compete with top thermoelectric materials such as SnSe whose maximum $ZT$ is $> 2.6$ \cite{Zhao2014}.
In addition, studies were mainly conducted on conventional halide perovskites, while emerging low-dimensional metal halides have yet to be explored.
Recently, we reported a high thermoelectric potential in metal halide \ce{Cs3Cu2I5} for the first time where asymmetric heat and charge transport in the material enables a high maximum $ZT$ of $2.6$ at $600$\,K \cite{Jung2021}.

In this article, we present \ce{CsCu2I3} as a candidate for thermoelectric applications on the basis of first-principles predictions.
\ce{CsCu2I3} is one of the copper-based low-dimensional halide compounds where 1D \ce{[Cu2I3]^-} anionic chains are separated by \ce{Cs^+} cations.
By performing lattice dynamics simulations, we found that heat transport in the material is highly anisotropic where the $\kappa_{\rm latt}$ in the $ab$-plane (perpendicular to the chains) is about $2$ times lower than that along the $c$-axis (chain direction).
We also confirmed that the experimentally reported $Cmcm$ structure of the material is not dynamically stable but an average of $Amm2$ structures during our phonon analysis.
Interestingly, electron transport in the material shows an opposite anisotropy compared to the phonon transport; electron mobility in the $ab$-plane is $1.5$ times higher than that along the $c$-axis. 
Due to this unique anisotropy of heat and electron transport in a single material, we predict that the \ce{CsCu2I3} in $Amm2$ structure reaches a $ZT$ of 2.2 at $600$\,K along the $b$-axis, having potential as a next-generation thermoelectric material.


\section{Results and Discussion}

\subsection{First-principles thermoelectrics workflow}

%
\begin{figure*}
\centering
\includegraphics[width=0.65\textwidth]{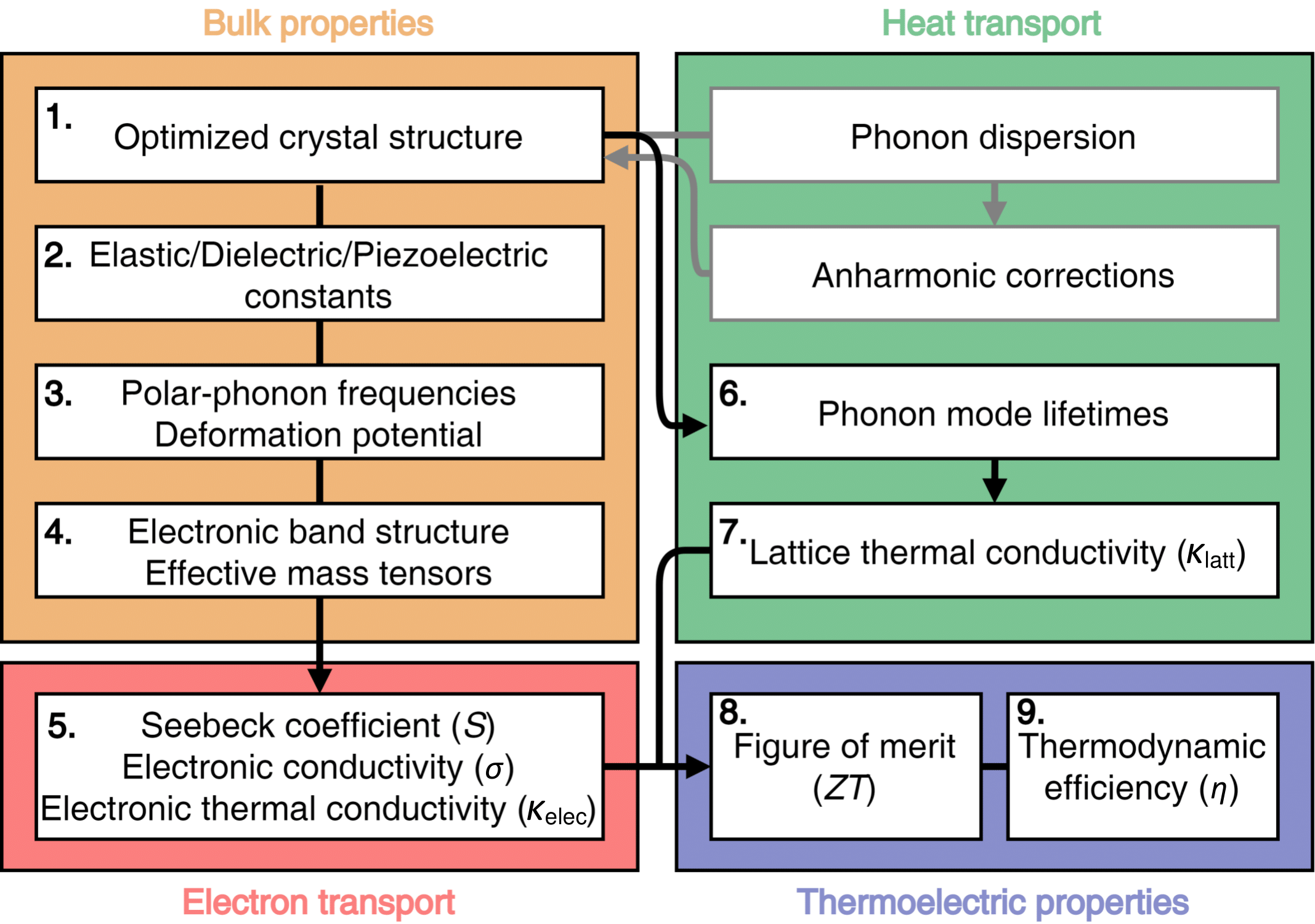}
\caption{\label{figure_1} Diagram for a first-principles thermoelectrics assessment workflow (total 9 steps). Steps which are optional or only required in certain cases are shown by gray boxes and arrows.}
\end{figure*}
The workflow to predict $ZT$ from first principles is shown in Fig.\,\ref{figure_1}.
In principle, this only requires prior knowledge of a crystal structure and all other properties can be directly calculated in turn. 
As the first step, the crystal structure of a compound of interest should be optimized to a local minimum in the potential energy surface. 
If the compound shows imaginary phonon modes, additional crystal structure optimization and/or anharmonic corrections should be applied to obtain dynamically stable structure, so that reliable thermal properties can be calculated \cite{Pallikara2022}.  
With the dynamically stable crystal structure, bulk properties can be assessed.
Results from steps 2--4 are given as inputs for calculating carrier lifetimes and transport properties -- $S$, $\sigma$, and $\kappa_{\rm elec}$ (step 5), while lattice thermal conductivity, $\kappa_{\rm latt}$ is obtained from anharmonic phonon calculations (steps 6 and 7).
Finally, thermoelectric properties such as $ZT$ and thermodynamic efficiency ($\eta$) are predicted by combining the outputs from steps 5 and 7.
Computational details are provided in section \ref{sec_methods}. 

\subsection{Structural analysis}

%
\begin{figure*}
\centering
\includegraphics[width=1.0\textwidth]{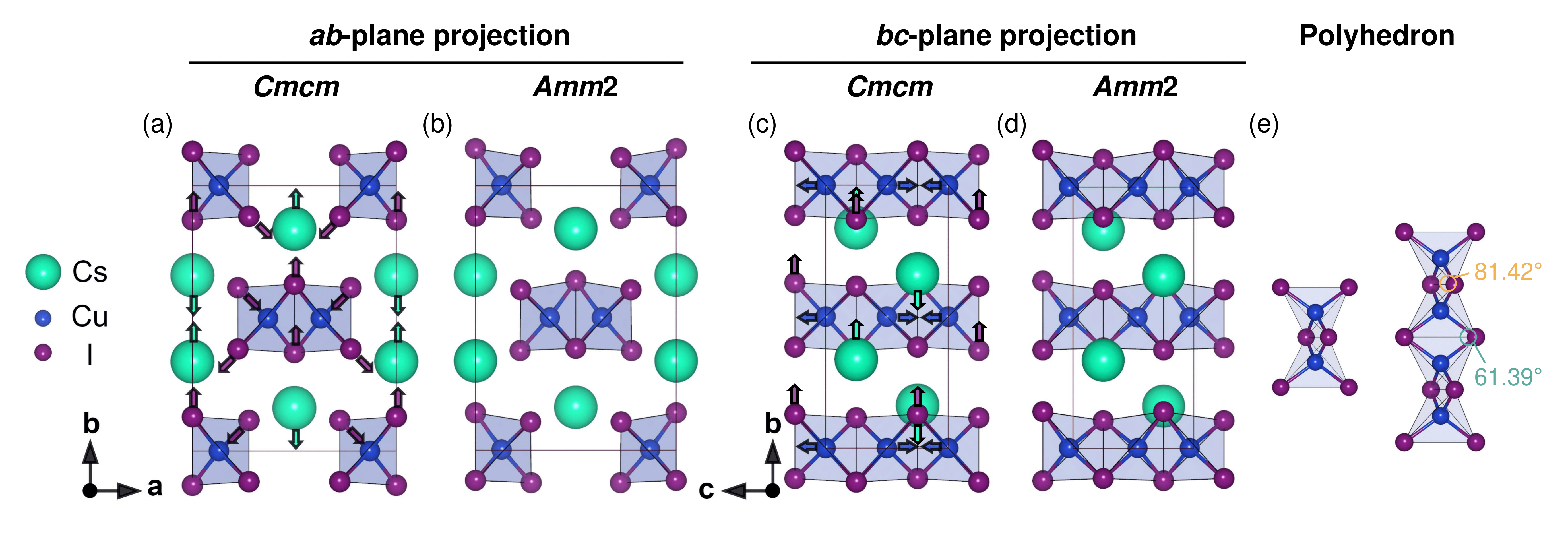}
\caption{\label{figure_2}Projected crystal structure of \ce{CsCu2I3} (a) $Cmcm$ and (b) $Amm2$ to the $ab$-plane, and to the $bc$-plane of (c) $Cmcm$ and (d) $Amm2$.  Arrows indicate the directions which the atoms move when the $Cmcm$ to $Amm2$ transition occurs. (e) Polyhedron of $Amm2$ (left) and its linked structure (right). Cu--I--Cu bond angles are indicated.}
\end{figure*}

\ce{CsCu2I3} has been reported in a $Cmcm$ space group,\cite{Jouini1980, Jun2019} where 1D \ce{[Cu2I3]^-} anionic chains are separated by \ce{Cs^+} cations as shown in Fig.\,\ref{figure_2}. 
When the experimentally known structure was adopted and optimized, we found that imaginary phonon modes are persistent across the first Brillouin zone (see Fig.\,\ref{figure_3}(a)), which confirms dynamic structural instability of the $Cmcm$ structure. (This will be further discussed in section \ref{subsec:dynamic}). 
To obtain a dynamically stable crystal structure, we deformed the $Cmcm$ structure along the eigenvector of its imaginary $\Gamma$ phonon mode, which results in a structural transition to new lower-symmetry $Amm2$ phase.
The structural transition from $Cmcm$ to $Amm2$ occurs with atomic positions shifted along the directions indicated by the arrows in Fig.\,\ref{figure_2}(a), (c).
Cs and I atoms move within the $b$-axis and $ab$-plane, respectively, while Cu atoms move along the $c$-axis.
The Cu--I--Cu bond angle for $Cmcm$ is alternately $71.54$\,$^{\circ}$ and $71.16$\,$^{\circ}$, while polyhedra distortion in $Amm2$ results in a bond angle of $61.39$\,$^{\circ}$ and $81.42$\,$^{\circ}$, as shown in Fig.\,\ref{figure_2}(e).
In $Cmcm$, the Cu--I bond lengths in the \ce{[CuI4]$^{3-}$} tetrahedron are of a similar value ($2.62$ and $2.61$\,{\AA}, two each), as opposed to the bonds in $Amm2$ all having a different value ($2.66$, $2.65$, $2.58$, and $2.61$\,{\AA}).
Thus, while the crystal system is maintained as orthorhombic, the crystal symmetry is lowered from centrosymmetric $Cmcm$ to non-centrosymmetric $Amm2$.
Comparison of the calculated lattice parameters of $Cmcm$ and $Amm2$, as well as experimental value from X-ray diffraction measurement are provided in Table \ref{table_2}.
%
\begin{table}
\caption{\label{table_2}%
Calculated lattice parameters ($a_0$, $b_0$, $c_0$) and volume ($V_0$) of the conventional orthorhombic unit cell for both polymorphs of \ce{CsCu2I3}, and experimental value from X-ray diffraction measurement.
}
\begin{ruledtabular}
\begin{tabular}{ccccc}
              & $a_0$\,({\AA})    & $b_0$\,({\AA})     & $c_0$\,({\AA})    & $V_0$\,({\AA}$^3$)  \\
\hline
$Cmcm$ & 10.06 & 13.08 & 6.10 & 802.3 \\
$Amm2$ & 10.03 & 13.19 & 6.10 & 807.9 \\
Exp. \cite{Roccanova2019}    & 10.55 & 13.17 & 6.10 & 847.4 \\
\end{tabular}
\end{ruledtabular}
\end{table}
The structural transition results in $a_0$ decreasing $0.30$\,\% while $b_0$ increasing $0.84$\,\%, both changes due to the movement of Cs and I atoms.
In contrast, $c_0$ is equivalent in both structures, as the shift of Cu atoms even out macroscopically.
$Amm2$ has an expanded volume of $0.70$\,\% compared to $Cmcm$.
The elastic and dielectric tensors calculated from $Cmcm$ and $Amm2$ phases are provided in Table S1.

\subsection{Dynamic structural instability}
\label{subsec:dynamic}

\begin{figure*}
\centering
\includegraphics[width=0.85\textwidth]{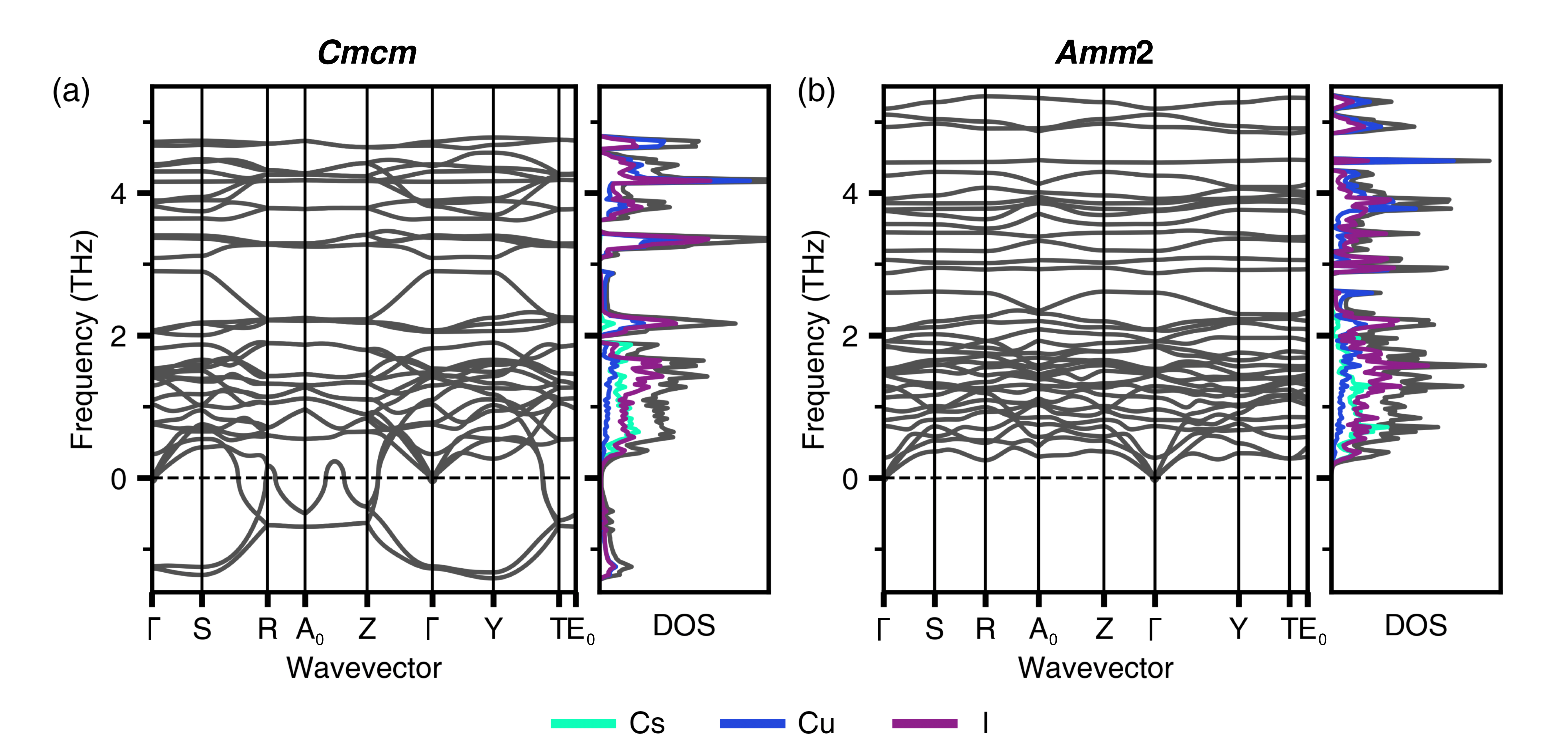}
\caption{\label{figure_3}Phonon dispersion of \ce{CsCu2I3}, (a) $Cmcm$ and (b) $Amm2$ structure. The atom-projected phonon density of states is plotted on the right side of the phonon dispersion.}
\end{figure*}

\begin{figure}
\centering
\includegraphics[width=0.37\textwidth]{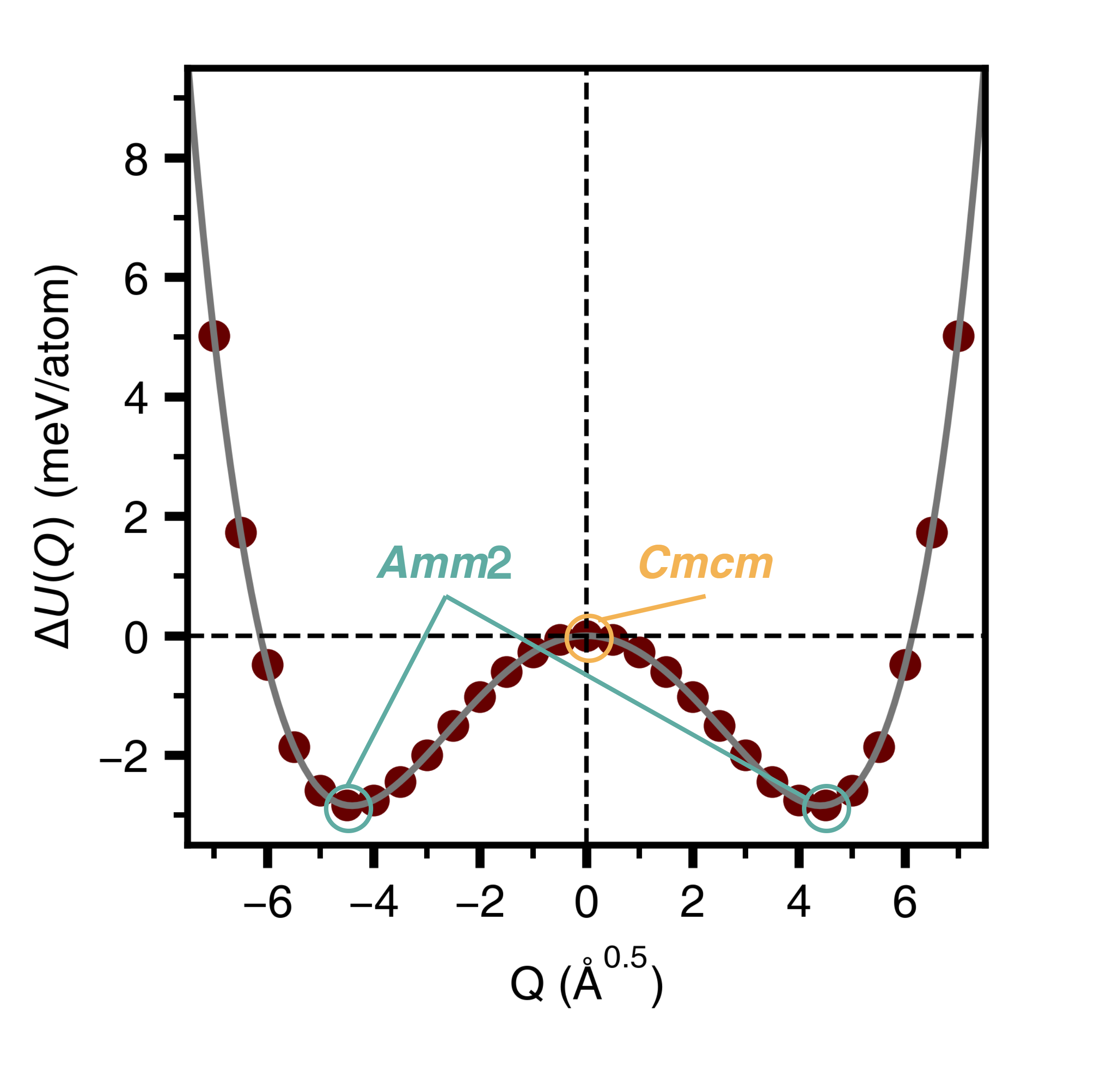}
\caption{\label{figure_4} Potential energy surface along the eigenvector components (arrows in Fig.\,\ref{figure_2}(a), (c)) for the imaginary mode at the $\Gamma$-point of \ce{CsCu2I3}, $Cmcm$ structure.  Filled circles represent calculated data points, and the solid line is a fit to a polynomial function.}
\end{figure}

The phonon dispersion of $Cmcm$ and $Amm2$ is illustrated together with the atom-projected phonon density of states (PDOS) in Fig.\,\ref{figure_3}(a), (b), respectively.
%
While the Materials Project \cite{Jain2013} repository, as well as computational\cite{PhysRevMaterials.4.095402} and experimental\cite{Roccanova2019} reports, indicate \ce{CsCu2I3} as a $Cmcm$ structure, dynamic structural instability of $Cmcm$ is evident by the numerous imaginary modes shown in Fig.\,\ref{figure_3}(a).
The eigenvector components for the imaginary mode at the $\Gamma$-point of $Cmcm$ are shown by arrows in Fig.\,\ref{figure_2}(a), (c).
Using \textsc{ModeMap}\cite{Skelton2016}, the corresponding energy as a function of the distortion amplitude along the eigenvectors illustrated in Fig.\,\ref{figure_2}(a), (c) is shown in Fig.\,\ref{figure_4}.
A characteristic double-well potential-energy curve is shown.
The saddle point corresponds to the $Cmcm$ structure, while the two wells indicate the lower symmetry $Amm2$ structure as a local minimum.
Thus, the energy-lowering distortion causes a structural transition to a ground-state polymorph, $Amm2$, with an energy $2.84$\,meV/atom lower than $Cmcm$.
As shown in Fig.\,\ref{figure_3}(b), the absence of imaginary phonon modes indicates the dynamic stability of $Amm2$.
Hence, we propose that the previously reported centrosymmetric $Cmcm$ structure is a macroscopic average over locally non-centrosymmetric $Amm2$ structures.
In addition, in the $Amm2$ structure, a lack of inversion symmetry causes a spontaneous electric polarization.
As shown in Fig.\,\ref{figure_2}(a), (c), polarization mainly occurs within the $ab$-plane by the shift of Cs and I atoms, while minute polarization along the $c$-axis corresponds to the movement of Cu atoms.
The corresponding piezoelectric tensor of the $Amm2$ structure is provided in Table S1.

%
%

\subsection{Ultra-low lattice thermal conductivity}

%
\begin{figure}
\centering
\includegraphics[width=0.37\textwidth]{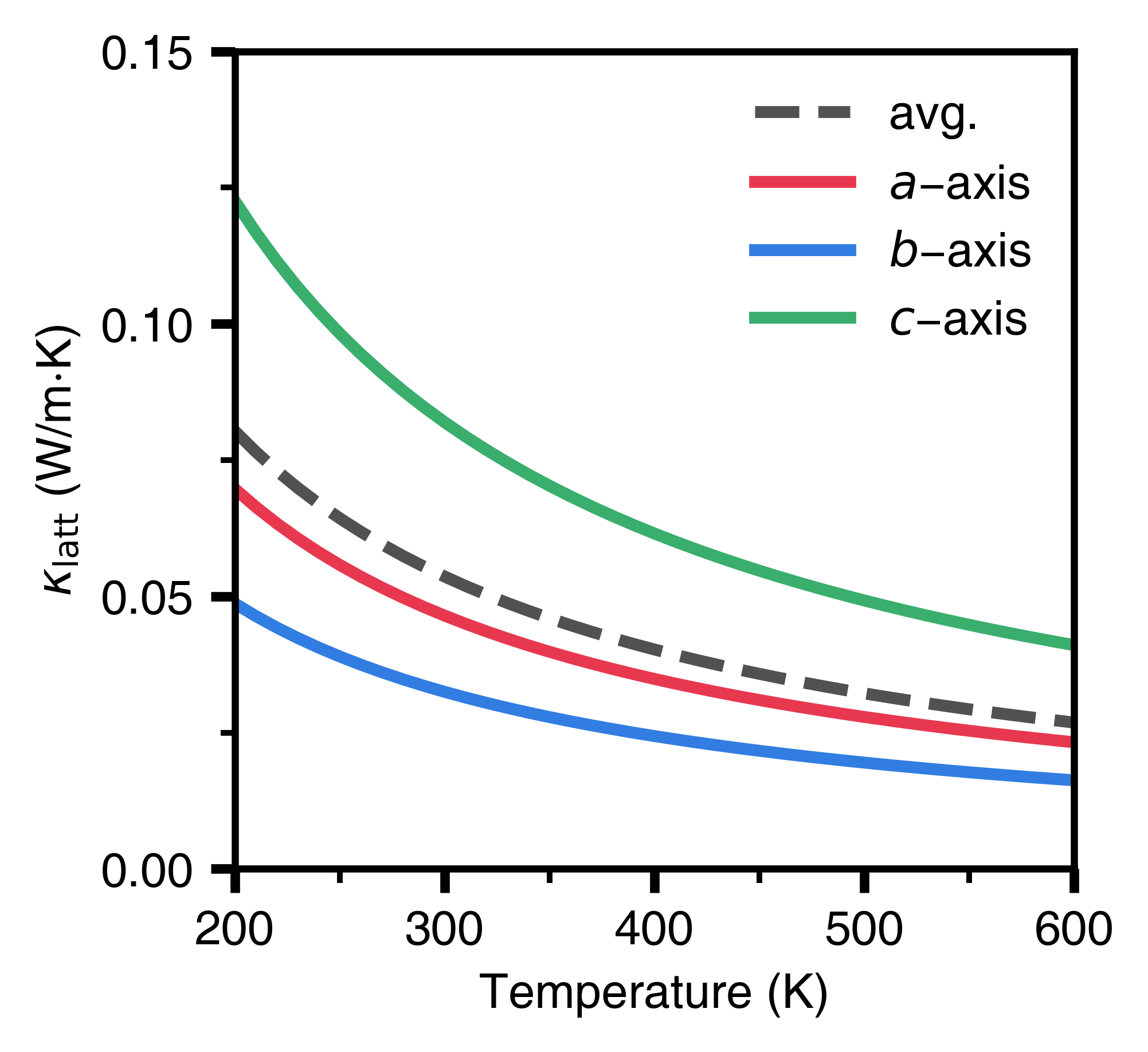}
\caption{\label{figure_5} Lattice thermal conductivity ($\kappa_{\rm latt}$) of \ce{CsCu2I3}, $Amm2$ structure, as a function of temperature along different crystallographic axes. `avg.' refers to the isotropically averaged $\kappa_{\rm latt}$ ($\kappa_{\rm avg}$) along the three ($a$-, $b$-, and $c$-) axes.}
\end{figure}

Fig.\,\ref{figure_5} shows the $\kappa_{\rm latt}$ of $Amm2$ \ce{CsCu2I3}, as a function of temperature along different crystallographic directions.
Because \ce{CsCu2I3} is experimentally reported to have a melting point at $\sim$644\,K \cite{Roccanova2019, Cheng2021}, the temperature range for calculating $\kappa_{\rm latt}$ as well as transport and thermoelectric properties discussed later is set up to $600$\,K.
$Amm2$ shows an unexpectedly low $\kappa_{\rm latt}$ (i.e. ultra-low $\kappa_{\rm latt}$) of a value under $0.1$\,W/m$\cdot$K for all directions even at $300$\,K; 0.05, 0.03, and $0.08$\,W/m$\cdot$K for the $a$-, $b$-, and $c$-axes, respectively.
The isotropically averaged $\kappa_{\rm latt}$ ($\kappa_{\rm avg}$) at $300$\,K is $0.05$\,W/m$\cdot$K, which is lower than one of the top thermoelectric materials, SnSe (0.2 W/m$\cdot$K at 300 K \cite{Sanchez2015}).
The value is slightly higher compared to \ce{Cs3Cu2I5} ($\kappa_{\rm avg}$ = $0.02$\,W/m$\cdot$K at room temperature (RT) \cite{Jung2021}), which was calculated at a similar level of theory.
At $600$\,K, the $\kappa_{\rm latt}$ values are $0.02$\,W/m$\cdot$K along the $a$- and $b$-axes, and $0.04$\,W/m$\cdot$K along the $c$-axis ($\kappa_{\rm avg}$ being $0.03$\,W/m$\cdot$K).
The anisotropy of \ce{CsCu2I3}, having a higher $\kappa_{\rm latt}$ along the $c$-axis, can be ascribed to a weaker chemical bonding towards the $c$-axis of the unit cell \cite{Ying2017}.
It is noteworthy that the $c$-axis is parallel to the \ce{[Cu2I3]$^{-}$} chains.

Acoustic phonon modes and low-frequency optic modes act as the primary heat carriers in crystals, mainly contributing to $\kappa_{\rm latt}$.
As shown in Fig.\,\ref{figure_3}(b), the low-lying optic modes of $Amm2$ are relatively flat, which leads to low group velocities ($v_{\lambda}$), one of the reasons for its ultra-low $\kappa_{\rm latt}$.
In addition, the high density of the low-lying optic modes produces a large number of scattering channels at this frequency range, causing short phonon lifetimes ($\tau_{\lambda}$).
Fig.\,S1 shows avoided crossings of the acoustic and low-frequency optic modes at the $\Gamma$--Y direction.
Avoided crossing is a characteristic feature shown when a `rattler' is present in the material \cite{Rahim2021}.
PDOS shown in Fig.\,\ref{figure_3}(b) indicates that lower-frequency phonon modes mostly comprise motions of Cs atoms.
Thus, we can infer that Cs atoms behave as the rattler, rattling within the space between \ce{[Cu2I3]$^-$} chains.
Fig.\,S2 shows the Cs--I bonds (total 10), and the broad range of bond lengths being from 3.81 to 4.21 {\AA} contributes to the anharmonicity of \ce{CsCu2I3}.
This is similar to the origin of anharmonicity of SnSe \cite{Zhao2016}.
We note that fluctuations between $Cmcm$ and $Amm2$ could also contribute to the scattering of the heat transport, but such higher-order anharmonicity is not considered here. 

%
\begin{figure*}
\centering
\includegraphics[width=1.0\textwidth]{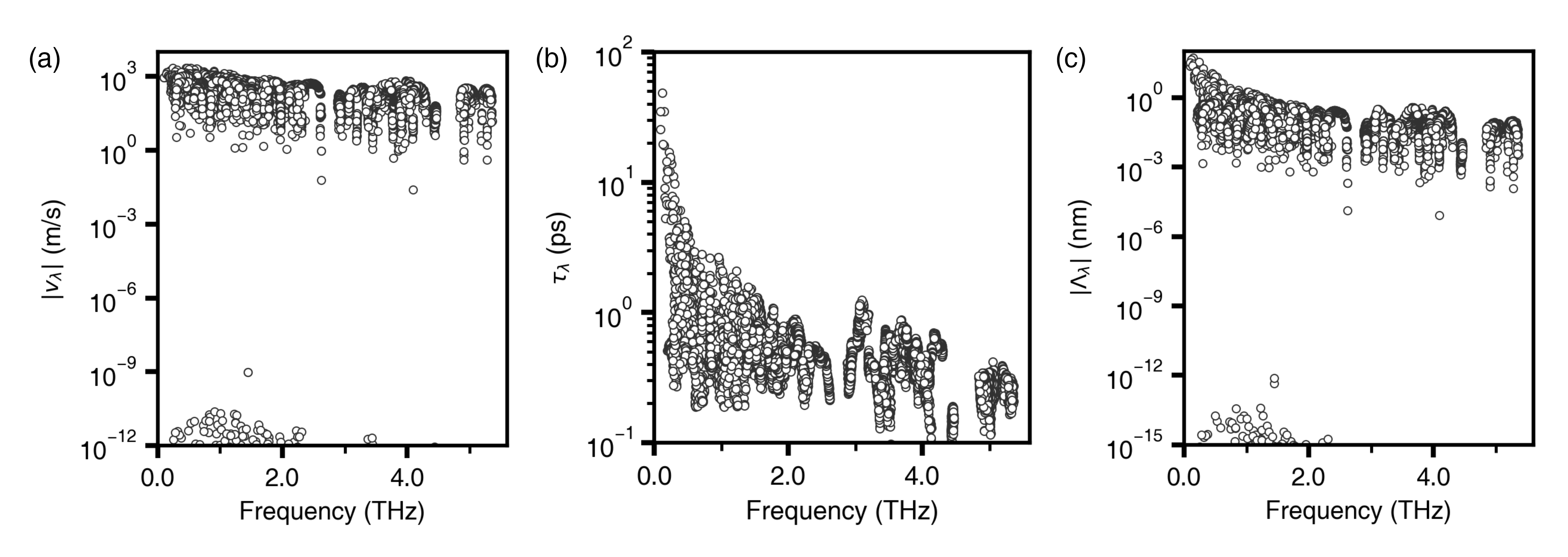}
\caption{\label{figure_6} Modal properties of the lattice thermal conductivity ($\kappa_{\rm latt}$) of \ce{CsCu2I3}, $Amm2$ structure, at $300$\,K; (a) group velocity norms ($v_{\lambda}$); (b) lifetimes ($\tau_{\lambda}$); and (c) mean free paths ($\Lambda_{\lambda}$ = $v_{\lambda}\times\tau_{\lambda}$).}
\end{figure*}

To further understand the origins of the ultra-low $\kappa_{\rm latt}$ of \ce{CsCu2I3}, we analyzed the modal contributions to the net transport (Fig.\,\ref{figure_6}).
The net transport, $\kappa_{\rm latt}$, is a sum of the individual phonon modes ($\lambda$):
\begin{equation}
 \kappa_{\rm latt} = \frac{1}{NV_0}\sum_{\lambda}\kappa_{\lambda} = \frac{1}{NV_0} \sum_{\lambda}C_{\lambda}v_{\lambda}\otimes v_{\lambda}\tau_{\lambda}
\label{latt_cond}
\end{equation}
where $N$ is the number of unit cells in the crystal (equivalent to the number of wavevectors included in the Brillouin zone summation), $V_0$ is the volume of the crystallographic unit cell, and $C_{\lambda}$ is the modal heat capacity.
The frequency spectra of $v_{\lambda}$, $\tau_{\lambda}$, and phonon mean free path ($\Lambda_{\lambda}$ = $v_{\lambda} \times \tau_{\lambda}$) at $300$\,K is shown in Fig.\,\ref{figure_6}(a)--(c), respectively \cite{phono3py_power_tools}.
In the entire frequency range, the majority of $v_{\lambda}$ fall within the range between $1$ and $10^3$\,m/s, and the fastest modes are seen at the 0--0.2\,THz frequency range.
The fastest modes correspond to the acoustic phonon bands that are relatively dispersive compared to the optic modes.
The overall spectra is comparable to \ce{(CH3NH3)PbI3} (\ce{MAPbI3}) \cite{Whalley2016}, a 3D perovskite reported to have a ultra-low $\kappa_{\rm latt}$ of $0.05$\,W/m$\cdot$K at $300$\,K, while the fastest modes have a lower $v_{\lambda}$ in \ce{CsCu2I3}.
In addition, a number of modes have a very low $v_{\lambda}$, from $10^{-12}$ to $10^{-10}$\,m/s, unseen in the $v_{\lambda}$ spectra of \ce{MAPbI3} \cite{Whalley2016} and \ce{Cs3Cu2I5} \cite{Jung2021}.
These modes correspond to the low-lying optic modes at the 0.2--2.4\,THz frequency range that are relatively flat.
The low $v_{\lambda}$ attribute to the heavy elements that constitute \ce{CsCu2I3}.

$\tau_{\lambda}$ mostly falls into the range of $10^{-1}$ to $10^{1}$\,ps, while a number of phonon modes within the 0--0.2\,THz frequency range (acoustic phonon modes) have a $\tau_{\lambda}$ longer than $10$\,ps.
The overall spectra is similar to \ce{MAPbI3} \cite{Whalley2016} and \ce{Cs3Cu2I5} \cite{Jung2021}, while the longest $\tau_{\lambda}$ of \ce{CsCu2I3} are longer compared to \ce{Cs3Cu2I5} (longest being $11$\,ps).
The combination of a low $v_{\lambda}$ and $\tau_{\lambda}$ leads to the majority of the modes having ${\Lambda}_{\lambda}$ shorter than $10^{0}$\,nm, which is why \ce{CsCu2I3} shows an ultra-low $\kappa_{\rm latt}$.
The low-frequency modes (acoustic and low-lying optic modes) have a relatively faster $v_{\lambda}$ and longer $\tau_{\lambda}$ resulting in a longer ${\Lambda}_{\lambda}$ compared to the high-frequency modes.
This matches with the fact that acoustic and low-lying optic modes are the primary heat carriers.

\subsection{Electronic structure and transport properties}

%
\begin{figure*}
\centering
\includegraphics[width=0.8\textwidth]{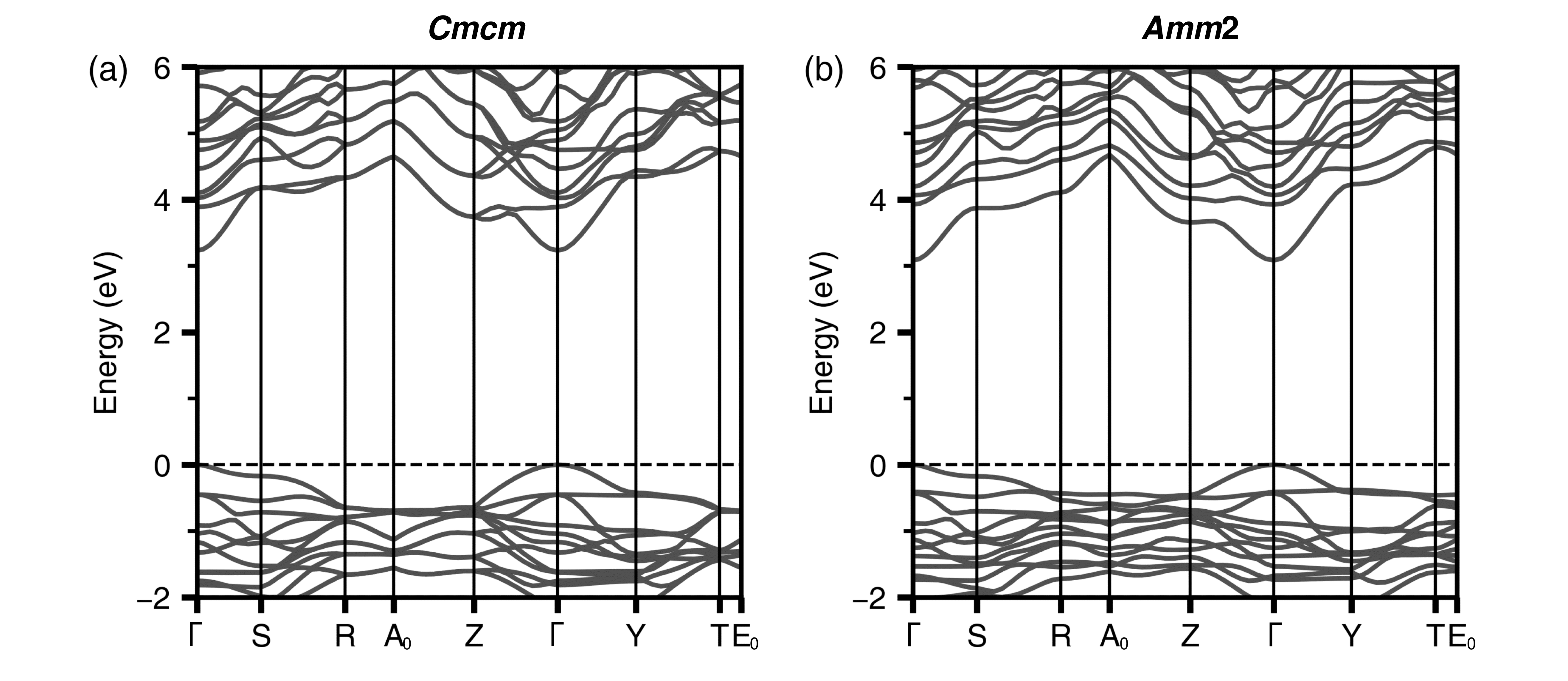}
\caption{\label{figure_7} Electronic band structure of \ce{CsCu2I3}, (a) $Cmcm$ and (b) $Amm2$ structure (the valence band maximum is set to $0$\,eV).}
\end{figure*}

The electronic band structure of $Cmcm$ and $Amm2$ is illustrated in Fig.\,\ref{figure_7}(a) and (b), respectively.
Both $Cmcm$ and $Amm2$ phases have a direct band gap ($E_g$) at the $\Gamma$-point, with a $E_g$ value of $3.23$\,eV and $3.08$\,eV, respectively.
The calculated $E_g$ is similar to the experimental value estimated from the optical absorption spectrum, $3.49$\,eV \cite{Cheng2021}.
As shown in Fig.\,S3, upper valence bands are dominated by Cu 3$d$ and I 5$p$ orbitals, while lower conduction bands arise from the hybridization of Cu 4$s$ and I 5$p$ orbitals.
For the conduction band minimum, I 5$s$ contributes more compared to I 5$p$.
Calculation of the orbitals that comprise band edges are in good agreement with previous reports \cite{Roccanova2019, Cheng2021}.
The contribution of Cs orbitals on those band edges is negligible, which is the well-known feature of low dimensional metal halides\cite{Jung2019, Lin2019}.
The corresponding orbitals are also equivalent to the electronic band structure of \ce{Cs3Cu2I5} \cite{Jung2021}.
The upper valence band is relatively flat, having a hole effective mass of $0.83$\,$m_e$ at the valence band maximum, while the lower conduction band is relatively dispersive with an electron effective mass of $0.31$\,$m_e$ at the conduction band minimum.
Conduction band has multiple valleys ($\Gamma$-point, Z-point, and along the S--R and Y--T direction), and the energy difference between the first and second conduction band edge is $0.57$\,eV for $Amm2$.
The dispersive nature and multiple valleys lead to a high $\sigma$ and $S$, respectively, suggesting the possibility of \ce{CsCu2I3} as a promising n-type thermoelectric material.

%
\begin{figure}
\centering
\includegraphics[width=0.37\textwidth]{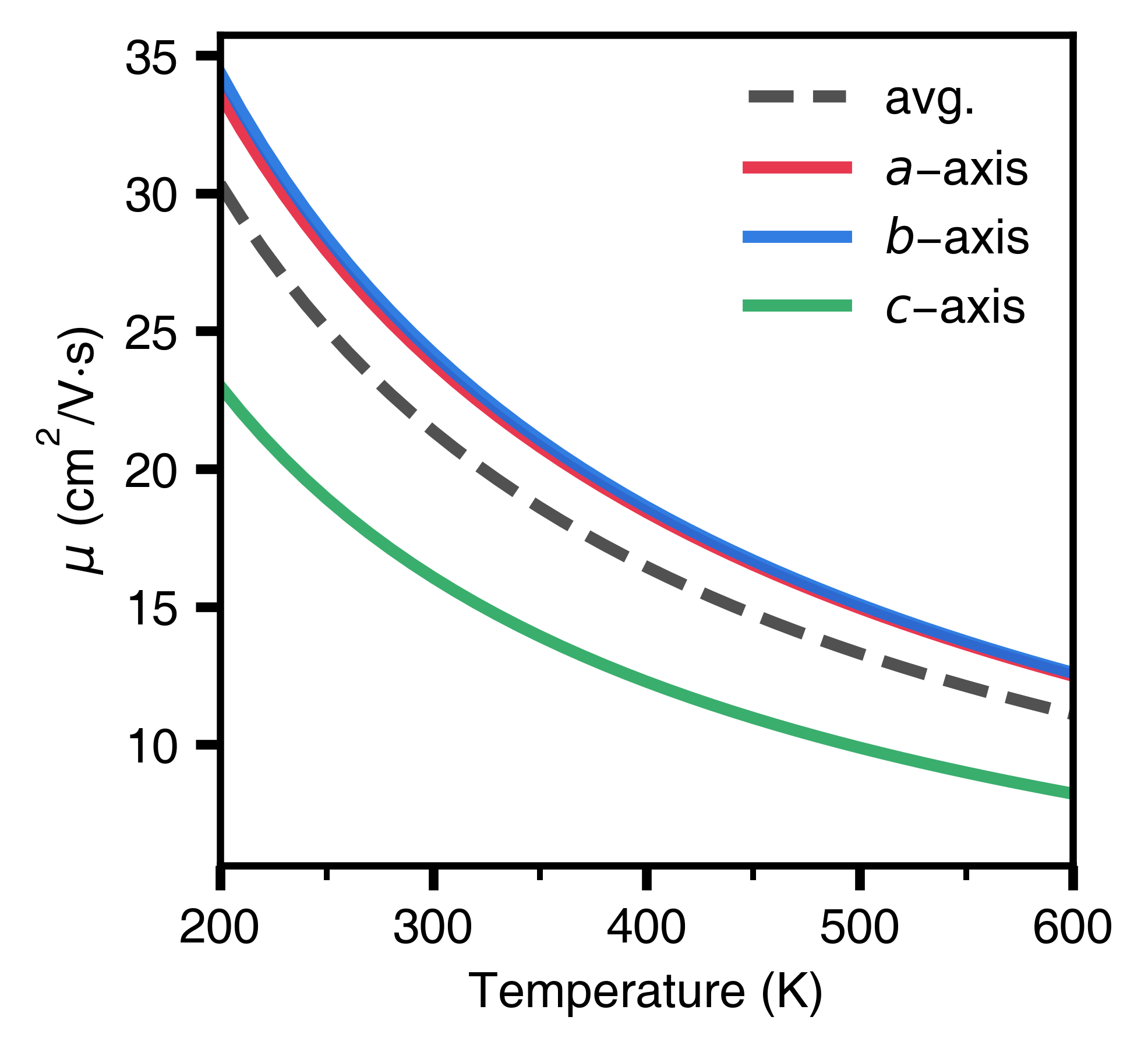}
\caption{\label{figure_8} Mobility ($\mu$) of \ce{CsCu2I3}, $Amm2$ structure, as a function of temperature along different crystallographic axes. `avg.' refers to the isotropically averaged $\mu$ along the three (a-, b-, and c-) axes (electron concentration = $6\times10^{18}$\,cm$^{-3}$).}
\end{figure}

Fig.\,\ref{figure_8} shows the calculated electron mobility, $\mu$, of $Amm2$ as a function of temperature along different crystallographic directions at the optimal electron concentration, $n_e$, ($6\times10^{18}$\,cm$^{-3}$) at which thermoelectric properties are maximized ($n_e$ will be further discussed in Fig.\,\ref{figure_10}(a)).
Similar to $\kappa_{\rm latt}$, $\mu$ is anisotropic, with $\mu$ being lower along the $c$-axis.
The isotropically averaged $\mu$ is $21.4$\,cm$^2$/V$\cdot$s at $300$\,K, which is slightly higher than the $\mu$ of \ce{Cs3Cu2I5} \cite{Jung2021} ($18.2$\,cm$^2$/V$\cdot$s at RT).
Fig.\,S4 shows $\mu$ of $Cmcm$ and $Amm2$ by the individual scattering mechanisms.
Acoustic deformation potential (ADP), ionized impurity (IMP), and polar optical phonon (POP) scattering mechanisms were considered for both structures, and for $Amm2$ (non-centrosymmetric), piezoelectric (PIE) scattering mechanism was considered as well.
$\mu$ is limited by POP scattering for both structures, followed by IMP and ADP scattering.
POP scattering is dominant in many of the top thermoelectric materials including SnSe \cite{Ma2018} and \ce{Cs3Cu2I5} \cite{Jung2021}.
In $Amm2$, PIE scattering has the smallest contribution to the total $\mu$, as its polarization is minute.

\begin{figure*}
\centering
\includegraphics[width=0.8\textwidth]{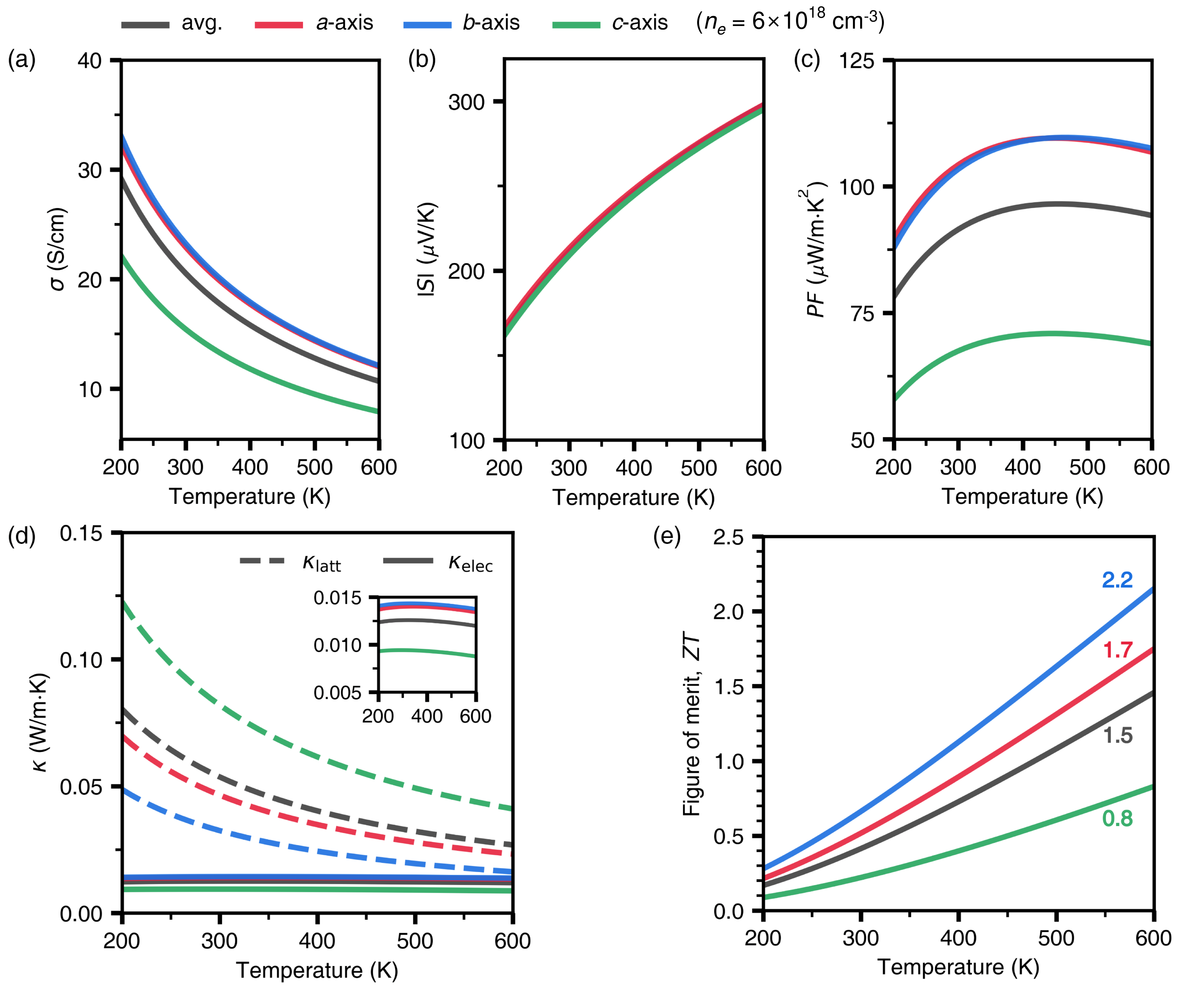}
\caption{\label{figure_9} Transport and thermoelectric properties of \ce{CsCu2I3} $Amm2$ structure, as a function of temperature along different crystallographic axes. (a) Electrical conductivity ($\sigma$); (b) Seebeck coefficient ($S$); (c) power factor ($PF$ = $S^2\sigma$); (d) electronic ($\kappa_{\rm elec}$, dashed lines) and lattice ($\kappa_{\rm latt}$, solid lines)  thermal conductivity (inset figure is a zoomed plot of $\kappa_{\rm elec}$); and (e) figure of merit ($ZT$) (electron concentration = $6\times10^{18}$\,cm$^{-3}$).}
\end{figure*}

The electronic transport properties -- $\sigma$, $S$, $PF$, and $\kappa_{\rm elec}$ -- of $Amm2$ as a function of temperature and $n_e$ are shown in Fig.\,S5.
$\sigma$ and $\kappa_{\rm elec}$ are proportional to $n_e$, but have an inverse relationship with temperature.
On the other hand, $\vert$S$\vert$ is disproportionate with $n_e$, and increases with temperature.
Fig.\,\ref{figure_9} shows $\sigma$, $S$, $PF$, and $\kappa_{\rm elec}$ as a function of temperature along different crystallographic directions ($n_e$ = $6\times10^{18}$\,cm$^{-3}$).
$\sigma$, $\kappa_{\rm elec}$, and $PF$ is higher along the $a$- and $b$-axes, which reflects the anisotropy of $\mu$.
$\vert$S$\vert$ is almost equivalent along all axes.
Along the $b$-axis, the $PF$ goes up to $109.66$\,$\mu$W/m$\cdot$K$^2$ at $470$\,K.

%
%

\subsection{Thermoelectric properties}

\begin{figure*}
\centering
\includegraphics[width=0.8\textwidth]{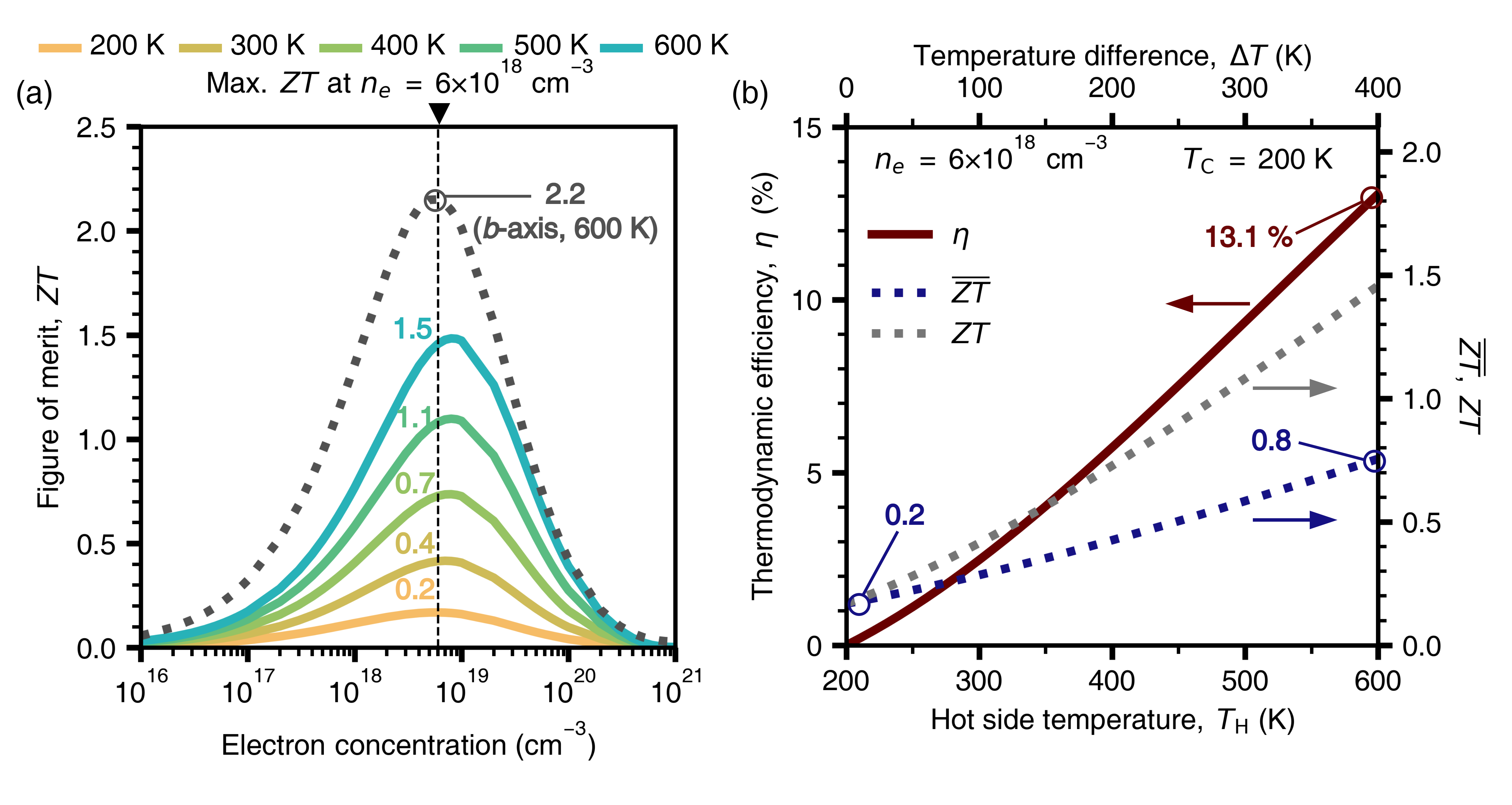}
\caption{\label{figure_10} (a) Calculated figure of merit ($ZT$) as a function of electron concentration and temperature. Isotropically averaged $ZT$ are given by solid lines, and $ZT$ along the $b$-axis (at $600$\,K) by the gray dotted line. (b) Thermodynamic efficiency ($\eta$, red), average $ZT$ ($\overline{ZT}$, blue), and $ZT$ (gray) as a function of the hot side temperature, $T_{\rm H}$, of \ce{CsCu2I3}, $Amm2$ structure.}
\end{figure*}

By combining the phonon and electron transport properties using Eq. (\ref{eqn_ZT}), $ZT$ for \ce{CsCu2I3} as a function of temperature along different crystallographic axes is predicted (Fig.\,\ref{figure_9}(e)).
Due to the anisotropy of $\sigma$, $\kappa_{\rm latt}$, and $\kappa_{\rm elec}$, $ZT$ is also anisotropic, showing a lower value along the $c$-axis.
At $600$\,K, it reaches a value of $2.2$ along the $b$-axis, while the $a$- and $c$-axes has a $ZT$ of $1.7$ and $0.8$, respectively.
Notably, a high $ZT$ is obtained at a lower temperature compared to the state-of-the-art thermoelectric material, n-type SnSe, which has a $ZT$ of $2.0$ above $700$\,K \cite{Wei2020}.
The origin of a high $ZT$ is a combination of ultra-low $\kappa_{\rm latt}$ and high $PF$.
Fig.\,\ref{figure_10}(a) shows the isotropically averaged $ZT$ of \ce{CsCu2I3} as a function of temperature and $n_e$.
$ZT$ is maximized at $n_e$ of $6\times10^{18}$\,cm$^{-3}$, and the maximum $ZT$ achievable at this condition is $1.5$ at $600$\,K.
As mentioned above, the highest $ZT$ from a conventional halide perovskite was only $0.15$, so this work may derive more attention towards \ce{CsCu2I3} and other low-dimensional metal halides.

%

Thermodynamic efficiency, $\eta$, of thermoelectric generators can be calculated by the following equation:
\begin{equation}
  \eta = \frac{T_{\rm H}-T_{\rm C}}{T_{\rm H}}\frac{\sqrt{1 + \overline{ZT}} - 1}{\sqrt{1 + \overline{ZT}} + \frac{T_{\rm C}}{T_{\rm H}}}
\label{eqn_efficiency}
\end{equation}
where $T_{\rm H}$ and $T_{\rm C}$ are the hot side and cold side temperature of the generator, respectively.
The average $ZT$ ($\overline{ZT}$) is defined as follows:
\begin{equation}
    \overline{ZT} = \frac{1}{(T_{\rm H}-T_{\rm C})}\int_{T_{\rm C}}^{T_{\rm H}}ZT dT 
\label{avgZT}
\end{equation}
Using these equations, we plotted $\eta$ as a function of $T_{\rm H}$, while $T_{\rm C}$ is fixed to $200$\,K (Fig.\,\ref{figure_10}(b)).
Maximum $\eta$ of $13.1$\,\% is achieved when the temperature difference is $400$\,K ($T_{\rm H}$ = $400$\,K), and $\overline{ZT}$ being 0.8.

The main concern when fabricating \ce{CsCu2I3} for thermoelectric applications would be whether the optimal $n_e$ could be achieved by doping.
Currently, doping of perovskite derivatives is underexplored, but below we address some of the doping strategies that can be implemented.
Similar to the doping approaches of 3D perovskites, doping in perovskite derivatives could be achieved either by (1) adding dopant sources to the precursor solution; (2) post-synthesis solution doping; and (3) post-synthesis vapor doping \cite{Amerling2021}.
The specific methods could be adding atomic dopants, and molecular dopants by surface doping.
In general, the B-site metals (Cu for \ce{CsCu2I3}) mainly contribute to the edge of the band structure \cite{Amerling2021} (c.f. Fig.\,S3).
Thus, substituting Cu$^+$ with +2 charged ions is more likely to produce the necessary states to tune $n_e$, compared to the doping of Cs and I.

$ZT$ of \ce{CsCu2I3} is relatively low compared to \ce{Cs3Cu2I5}.
However, it is reported that \ce{CsCu2I3} is more stable than \ce{Cs3Cu2I5} when dopants are added, and the unstable \ce{Cs3Cu2I5} decompose and form into \ce{CsCu2I3} \cite{Fan2021}.
Thus, although a higher optimal $n_e$ is required for \ce{CsCu2I3} ($6\times10^{18}$\,cm$^{-3}$, compared to $4\times10^{18}$\,cm$^{-3}$ for \ce{Cs3Cu2I5}), it may be achieved more easily. 
In addition, whether this high thermoelectric performance is shown only for \ce{CsCu2I3} and \ce{Cs3Cu2I5}, or from perovskite derivatives in general requires further investigation.
Through our initial calculations, copper-based metal halides \ce{K2CuX3} and \ce{CsCu2X3} (X = Cl, Br) are also expected to have a high potential as an n-type thermoelectric material.
Changing the A-site or X-site elements to the ones in the same group of the periodic table is also worth a try.
Thus, investigating \ce{Cs3Cu2X5} and \ce{Rb2CuX3} (X = Cl, Br) as well as \ce{RbCu2Y3} (Y = Br, I) could be possible.

\section{Conclusions}

In this paper, we report a new direction for metal halide thermoelectrics with a predictive study on the structure, properties, and performance of \ce{CsCu2I3}.

The dynamic structural instability of the previously known $Cmcm$ structure of \ce{CsCu2I3} was investigated.
We report a new, ground-state $Amm2$ structure of \ce{CsCu2I3}, and compared its basic bulk properties with $Cmcm$.
The ultra-low $\kappa_{\rm latt}$ of $Amm2$ and its origins were studied in detail.
Additionally, the electronic transport properties as well as $ZT$ were first reported in this work.
We predict that \ce{CsCu2I3} is a new promising n-type  thermoelectric material, and require further investigations in low-dimensional metal halides.

The centrosymmetric $Cmcm$ structure is a macroscopic average over locally non-centrosymmetric $Amm2$ structures.
The octahedra distortion leads to an energy-lowering structural transition from $Cmcm$ to $Amm2$, and the energy being 2.84 meV/atom lower.
The lack of inversion symmetry in $Amm2$ results in a spontaneous lattice polarization mainly within the $ab$-plane.

$Amm2$ shows an ultra-low $\kappa_{\rm latt}$ with $\kappa_{\rm avg}$ at $300$\,K being $0.05$\,W/m$\cdot$K, and the values being higher along the $c$-axis (i.e. anisotropic).
The low $v_{\lambda}$ is due to the low-frequency optic modes being relatively flat.
Avoided crossings of the acoustic and low-lying optic modes are shown from the dispersion, which is the cause of short $\tau_{\lambda}$.
Cs atoms between the \ce{[Cu2I3]$^-$} chains behave as rattlers, and the inequivalent Cs--I bond lengths give rise to a strong anharmonicity.
The structural transition between $Cmcm$ and $Amm2$ could also contribute to the phonon scattering.

The conduction bands of \ce{CsCu2I3} is relatively dispersive and has multiple valleys, which is the reason for its high $\sigma$ and $S$, respectively, characteristics of a novel n-type thermoelectric material.
POP is the dominant scattering mechanism for both $Cmcm$ and $Amm2$, and PIE scattering is also considered in $Amm2$ because of its lack of inversion symmetry.
Similar to $\kappa_{\rm latt}$, the electronic properties are also anisotropic (superior along the $a$- and $b$-axes).

The predicted $ZT$ of \ce{CsCu2I3} reaches $2.2$ at $600$\,K along the $b$-axis ($n_e$ = $6\times10^{18}$\,cm$^{-3}$), comparable to the $ZT$ of state-of-the-art  thermoelectric materials.
The origin of high $ZT$ is a combination of ultra-low $\kappa_{\rm latt}$ and high $PF$.
$\eta$ of $13.1$\,\% is achievable when \ce{CsCu2I3} is used in a thermoelectric generator ($T_{\rm H}$ = $600$\,K, $T_{\rm C}$ = $200$\,K).

\section{\label{sec_methods}Methods}

\subsection{\label{app:subsec} Density functional theory calculations}
Calculations of the total energy, electronic band structure, and inputs for the AMSET package such as the dielectric, elastic, and piezoelectric constants were performed using density functional theory (DFT) within periodic boundary conditions through the Vienna $Ab$ $Initio$ Simulation Package (VASP) \cite{Kresse1996a,Kresse1996b}.
Projector Augmented-Wave (PAW) \cite{Kresse1999c,Blochl1994} method was employed to explicitly treat the valence state of Cs, Cu, and I atoms as $9~(5s^{2}5p^{6}6s^{1})$, $17~(2p^{6}3d^{10}4s^{1})$, and $7~(5s^{2}5p^{5})$ electrons, respectively.

For structure optimization, the Perdew-Burke-Ernzerhof exchange-correlation (xc) functional revised for solids (PBEsol) \cite{Perdew2008} was used with a $6\times6\times8$ $\Gamma$-centered $k$-mesh, a plane-wave kinetic energy cutoff of $700$\,eV, and the convergence criteria set to $10^{-8}$\,eV and $10^{-4}$\,eV/{\AA} for the total energy and atomic forces, respectively.
The elastic and dielectric constant was calculated using the finite-displacement (FD) method and density functional perturbation theory (DFPT), respectively.
The bulk modulus was calculated using the Phonopy \cite{Togo2015a} code by fitting the energy-volume to the third-order Birch-Murnaghan equation of state \cite{PhysRev.71.809}.

Calculations of the electronic band structure and electron transport were done using the hybrid DFT functional of Heyd, Scuseria, and Ernzerhof (HSE06) \cite{Heyd2003}.
Compared to the structure optimization, a denser $k$-mesh of $12\times12\times16$ was used, while the kinetic energy cutoff was lowered to $400$\,eV.
The hole and electron effective mass, $m^{*}$, was calculated using the sumo \cite{Ganose2018} code, which uses parabolic fitting by the following equation:
\begin{equation}
\frac{1}{m^{*}} = \frac{\partial^2E(k) }{\partial k^2}\frac{1}{\hbar^{2}}
\label{eff_mass}
\end{equation}
where $E(k)$ is the band energy as a function of the electron wavevector $k$, and $\hbar$ is the reduced Plank's constant.
The electronic band structure calculated above was used as the input. 

\subsection{Structure distortion}
Harmonic level phonon calculations were performed using the Phonopy \cite{Togo2015a} code with VASP as the force calculator.
The second-order interatomic force constants (IFCs) were computed using the supercell FD approach with a $3\times3\times3$ $k$-mesh, step size of $0.01$\,{\AA}. A total of $11$ displacements for $Cmcm$ and $22$ displacements for $Amm2$ were calculated.
A $2\times2\times3$ supercell of the $12$\,-atom unit cell ($144$\, atoms), was employed for both structures.
The ModeMap \cite{Skelton2016} code was used to compute the displacement of the atoms, $u_{j,l}$ ($j$th atom in the $l$th unit cell), along an imaginary-mode eigenvector, $W_{\lambda, j}$ ($\lambda$ is the phonon mode), at the $\Gamma$-point:
\begin{equation}
u_{j,l}= \frac{1}{\sqrt{n_am_j}}Re\left [ \sum_{\lambda}Q_{\lambda}W_{\lambda, j}e^{-iq\cdot r_{j,l}} \right ]
\label{modemap}
\end{equation}
where $m_j$ is the atomic mass, $n_a$ is the number of atoms in the supercell used to model the displacement, $Q_{\lambda}$ is the distortion amplitude, $q$ is the phonon wavevector, and $r_{j,l}$ is the atomic position.
Post processing was also performed using the code to map the energy, ${\Delta}U(Q)$, as a function of $Q_{\lambda}$ along the given $W_{\lambda, j}$ (c.f. Fig.\,\ref{figure_4}).
The ground-state structure, $Amm2$, was then obtained using the structure at the energy minimum.

\subsection{Phonon and electron transport}
$\kappa_{\rm latt}$ calculations were carried out using the Phono3py \cite{Togo2015b} code, solving the linearized Boltzmann transport equation (BTE) using the single-mode relaxation-time approximation (RTA) (Eq. (\ref{latt_cond})).
The third-order IFCs were calculated with a FD step size of $0.03$\,{\AA}, and a total of $5568$\, displacements were considered in a $48$\,-atom unit cell.
A $q$-mesh of $12\times12\times16$ was employed to compute the lattice thermal conductivity.
Graphical analysis of the modal properties were performed using the Phonopy-power-tool \cite{phono3py_power_tools} code.
Convergence tests for the lattice thermal conductivity over $q$-mesh, and distribution of force norms for the force sets can be found in Fig.\,S6.

Unlike the BoltzTraP \cite{Madsen2006} code, the AMSET \cite{Ganose2021} package uses DFT band structures to solve the BTE without the constant RTA. The characteristic scattering rate, $\tau_e$, is calculated using the Matthiessen's rule:
\begin{equation}
  \frac{1}{\tau_e} = 
  \frac{1}{\tau^{\rm ADP}} + \frac{1}{\tau^{\rm IMP}} + \frac{1}{\tau^{\rm POP}} + \frac{1}{\tau^{\rm PIE}}
\label{eqn_matthies}
\end{equation}
The mode dependent scattering rates, from state $|n\mathbf{k}\rangle$ to state $|m\mathbf{k+q}\rangle$, is calculated using Fermi's golden rule:
\begin{equation}
    \tilde{\tau}_{n\mathbf{k}\rightarrow m\mathbf{k}+\mathbf{q}}^{-1} = 
        \frac{2\pi}{\hbar} \lvert g_{nm}(\mathbf{k}, \mathbf{q}) \rvert^2
        \delta(\varepsilon_{n\mathbf{k}} - 
        \varepsilon_{m\mathbf{k}+\mathbf{q}})
\label{eqn_Fermi_golden}
\end{equation}
where $\varepsilon$ is the electron energy, $\delta$ is the Dirac delta function and $g$ is the coupling matrix element.
The electron transport properties were computed by the generalized transport coefficients:
\begin{equation}
L_{\alpha \beta }^{n} = e^{2}\int \sum_{\alpha \beta}(\varepsilon)(\varepsilon-\varepsilon_{\rm F})^n\left [ -\frac{\partial f^0}{\partial \varepsilon} \right ]d\varepsilon
\label{eqn_gen_transport}
\end{equation}
where $\alpha$ and $\beta$ denotes Cartesian coordinates, $\Sigma_{\alpha \beta}(\varepsilon)$ is the spectral conductivity, $\varepsilon_{\rm F}$ is the fermi level at a certain doping concentration and temperature, and $f^0$ is the Fermi-Dirac distribution. The properties are obtained as
\begin{equation}
\sigma_{\alpha \beta}  = L_{\alpha \beta}^{0}
\label{eqn_elec_cond}
\end{equation}
\begin{equation}
S_{\alpha \beta}  =\frac{1}{eT} \frac{L_{\alpha \beta}^{1}}{L_{\alpha \beta}^{0}}
\label{eqn_seebeck}
\end{equation}
\begin{equation}
\kappa _{\alpha \beta}  =\frac{1}{e^2T}\left [\frac{{(L_{\alpha \beta}^{1})}^2}{L_{\alpha \beta}^{0}}-L_{\alpha \beta}^{2} \right] 
\label{eqn_elec_therm}
\end{equation}
As mentioned above, the required material parameters such as the dielectric, elastic, and piezoelectric constants, phonon frequencies, and deformation potential were determined by DFT calculations (Table S1).
As the valence bands are relatively flat, calculations were only conducted under n-type doping conditions, in the doping range from $10^{16}$ to $10^{21}$, and the temperature range from $200$\,K to $600$\,K.
The interpolation factor was set to 10 for all AMSET calculations.
Convergence tests for the electron transport calculations over $k$-mesh and the interpolation factor can be found in Fig.\,S5.

\begin{acknowledgments}
This work was supported by a National Research Foundation of Korea (NRF) grant funded by the Korean government (MSIT) (No. 2018R1C1B6008728).
Via membership of the UK’s HEC Materials Chemistry Consortium, which is funded by EPSRC (EP/L000202), this work used the ARCHER2 UK National Supercomputing Service (http://www.archer2.ac.uk).

The authors declare that they have no competing financial interests or personal relationships that could have influenced the work reported in this paper. 

An online repository containing the basic bulk properties, force constant sets,
and raw AMSET input/output files have been made available at https:doi.org/10.xxx/zenodo.xxx. [link will be added later].

J.W.P. performed the calculations and data analysis, and wrote the original draft under the supervision of Y.-K.J. and A.W. 
All authors contributed to discussing the results.
\end{acknowledgments}




\bibliography{manuscript}

\end{document}



\title{Supplemental Material: \\Metal halide thermoelectrics: prediction of high-performance \ce{CsCu2I3}}

\author{Jong Woong Park}
\affiliation{
Department of Materials Science and Engineering, Yonsei University, Seoul 03722, South Korea}%
\author{Young-Kwang Jung}
 \email{yj359@cam.ac.uk; a.walsh@imperial.ac.uk}
 \affiliation{
Department of Chemical Engineering and Biotechnology, University of Cambridge, Cambridge CB3 0AS, UK}%
\author{Aron Walsh}
\email{yj359@cam.ac.uk; a.walsh@imperial.ac.uk}
\affiliation{Department of Materials, Imperial College London, Exhibition Road, London SW7 2AZ, UK}%

\maketitle


\renewcommand{\thetable}{S\arabic{table}} 
\renewcommand{\thefigure}{S\arabic{figure}}

%
\begin{table*}[h]
\caption{Calculated basic bulk properties of \ce{CsCu2I3}, $Cmcm$ and $Amm2$ structure, used as inputs of the AMSET package.}
\label{table_S1}
\begin{ruledtabular}
\begin{tabular}{ccccccccccccc}
                             & \multicolumn{6}{c}{\textit{Cmcm}}                                                & \multicolumn{6}{c}{$Amm2$}                                                \\
\hline
Elastic const. (GPa)                             & 22.37        & 9.73       & 16.97        & 0.00       & 0.00        & 0.00      \noalign{\hskip 2mm} & 22.10        & 9.85       & 15.91        & 0.00       & 0.00        & 0.00       \\
\noalign{\vskip 1mm}
\textit{}                  [$6\times6$]  & 9.73         & 18.10      & 9.15        & 0.00       & 0.00        & 0.00      \noalign{\hskip 2mm} & 9.85         & 17.33      & 7.68        & 0.00       & 0.00        & 0.00       \\
\noalign{\vskip 1mm}
            & 16.97        & 9.15      & 20.11        & 0.00       & 0.00        & 0.00      \noalign{\hskip 2mm} & 15.91        & 7.68      & 29.81        & 0.00       & 0.00        & 0.00       \\
\noalign{\vskip 1mm}
                         & 0.00         & 0.00       & 0.00         & 5.68       & 0.00        & 0.00      \noalign{\hskip 2mm} & 0.00         & 0.00       & 0.00         & 6.87       & 0.00        & 0.00       \\
\noalign{\vskip 1mm}
                             & 0.00         & 0.00       & 0.00         & 0.00       & 5.02       & 0.00      \noalign{\hskip 2mm} & 0.00         & 0.00       & 0.00         & 0.00       & 4.54       & 0.00       \\
\noalign{\vskip 1mm}
                             & 0.00         & 0.00       & 0.00         & 0.00       & 0.00        & 13.93      \noalign{\hskip 2mm} & 0.00         & 0.00       & 0.00         & 0.00       & 0.00        & 13.18       \\
\noalign{\vskip 3mm}
Bulk modulus (GPa)                 & \multicolumn{6}{c}{13.39}                                                        \noalign{\hskip 2mm}& \multicolumn{6}{c}{13.31}                                                        \\
\noalign{\vskip 3mm}
High freq. dielectric const.                             & \multicolumn{2}{c}{4.57}  & \multicolumn{2}{c}{0.00}  & \multicolumn{2}{c}{0.00} & \multicolumn{2}{c}{4.57}  & \multicolumn{2}{c}{0.00}  & \multicolumn{2}{c}{0.00} \\
\noalign{\vskip 1mm}
[$3\times3$] & \multicolumn{2}{c}{0.00}  & \multicolumn{2}{c}{4.53}  & \multicolumn{2}{c}{0.00} & \multicolumn{2}{c}{0.00}  & \multicolumn{2}{c}{4.57}  & \multicolumn{2}{c}{0.00} \\
\noalign{\vskip 1mm}
                            & \multicolumn{2}{c}{0.00}  & \multicolumn{2}{c}{0.00}  & \multicolumn{2}{c}{4.88} & \multicolumn{2}{c}{0.00}  & \multicolumn{2}{c}{0.00}  & \multicolumn{2}{c}{4.87} \\
\noalign{\vskip 3mm}
Static dielectric const.                             & \multicolumn{2}{c}{8.70} & \multicolumn{2}{c}{0.00} & \multicolumn{2}{c}{0.00} & \multicolumn{2}{c}{9.48} & \multicolumn{2}{c}{0.00} & \multicolumn{2}{c}{0.00} \\
\noalign{\vskip 1mm}
[$3\times3$]     & \multicolumn{2}{c}{0.00} & \multicolumn{2}{c}{10.84}   & \multicolumn{2}{c}{0.00} & \multicolumn{2}{c}{0.00} & \multicolumn{2}{c}{11.04}   & \multicolumn{2}{c}{0.00} \\
\noalign{\vskip 1mm}
                         & \multicolumn{2}{c}{0.00}  & \multicolumn{2}{c}{0.00}  & \multicolumn{2}{c}{7.44} & \multicolumn{2}{c}{0.00}  & \multicolumn{2}{c}{0.00}  & \multicolumn{2}{c}{7.51} \\
\noalign{\vskip 3mm}
Piezoelectric const. (C/m$^2$)         & \multicolumn{6}{c}{}                                                             & 0.00        & 0.00      & 0.00         & -0.06      & 0.00        & 0.00       \\
\noalign{\vskip 1mm}
        [$3\times6$] & \multicolumn{6}{c}{-}                                                            & 0.04         & 0.05       & -0.07        & 0.00       & 0.00        & 0.00       \\
\noalign{\vskip 1mm}
                          & \multicolumn{6}{c}{}                                                             & 0.00         & 0.00       & 0.00         & 0.00       & -0.02       & 0.00       \\
\noalign{\vskip 3mm}
POP freq. (THz)              & \multicolumn{6}{c}{1.07}                                                         & \multicolumn{6}{c}{2.03}                                                         \\
\end{tabular}%
\end{ruledtabular}
\end{table*}
%


%
\begin{figure*}[h]
  \includegraphics[width=0.8\textwidth]{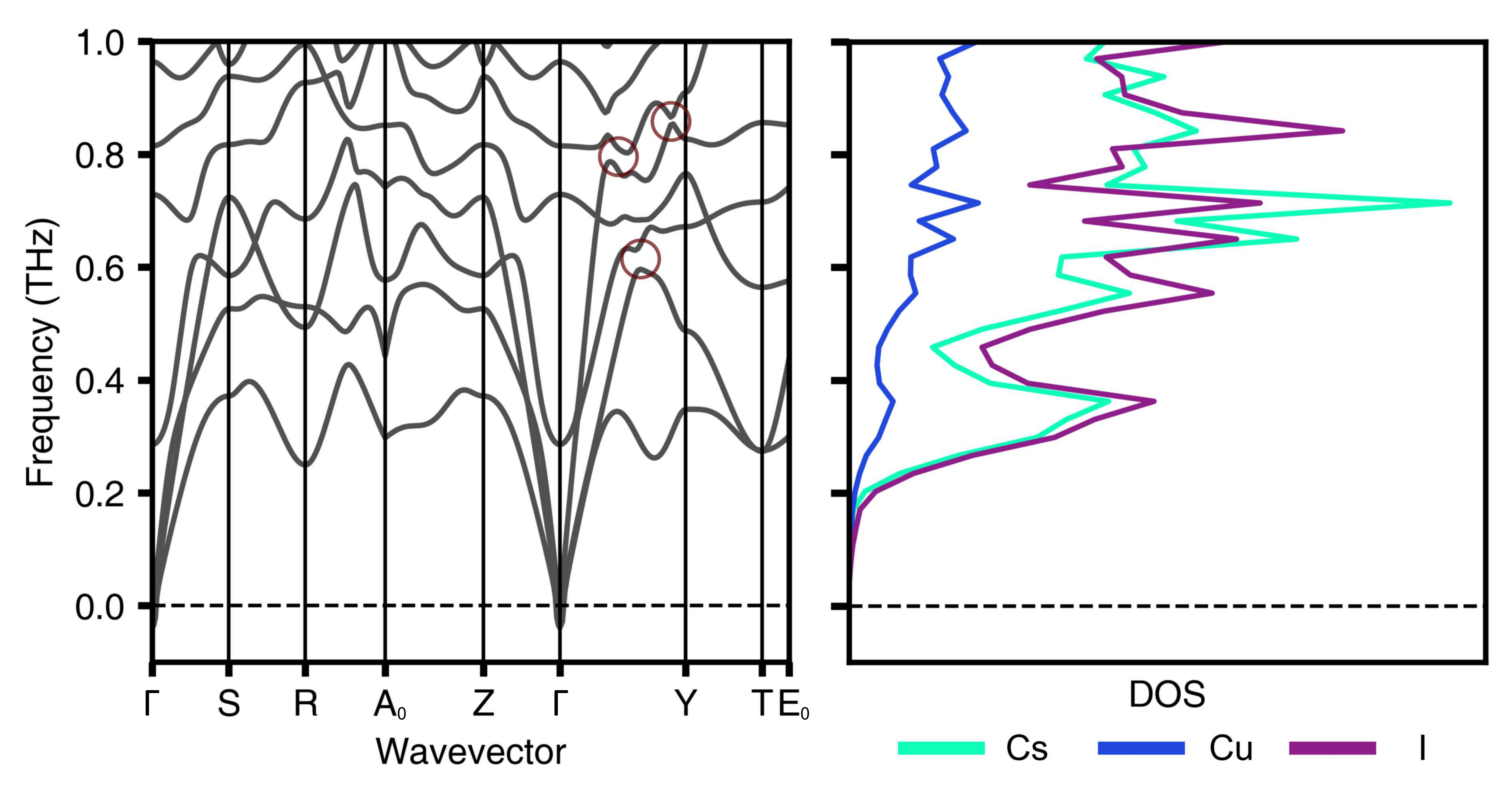}
  \caption{Zoomed phonon dispersion (left), which shows avoided crossings (empty circles), and atom-projected phonon density of states (right) of \ce{CsCu2I3}, $Amm2$ structure.}
  \label{figure_S1}
\end{figure*}
%

%
\begin{figure*}[h]
  \includegraphics[width=1.0\textwidth]{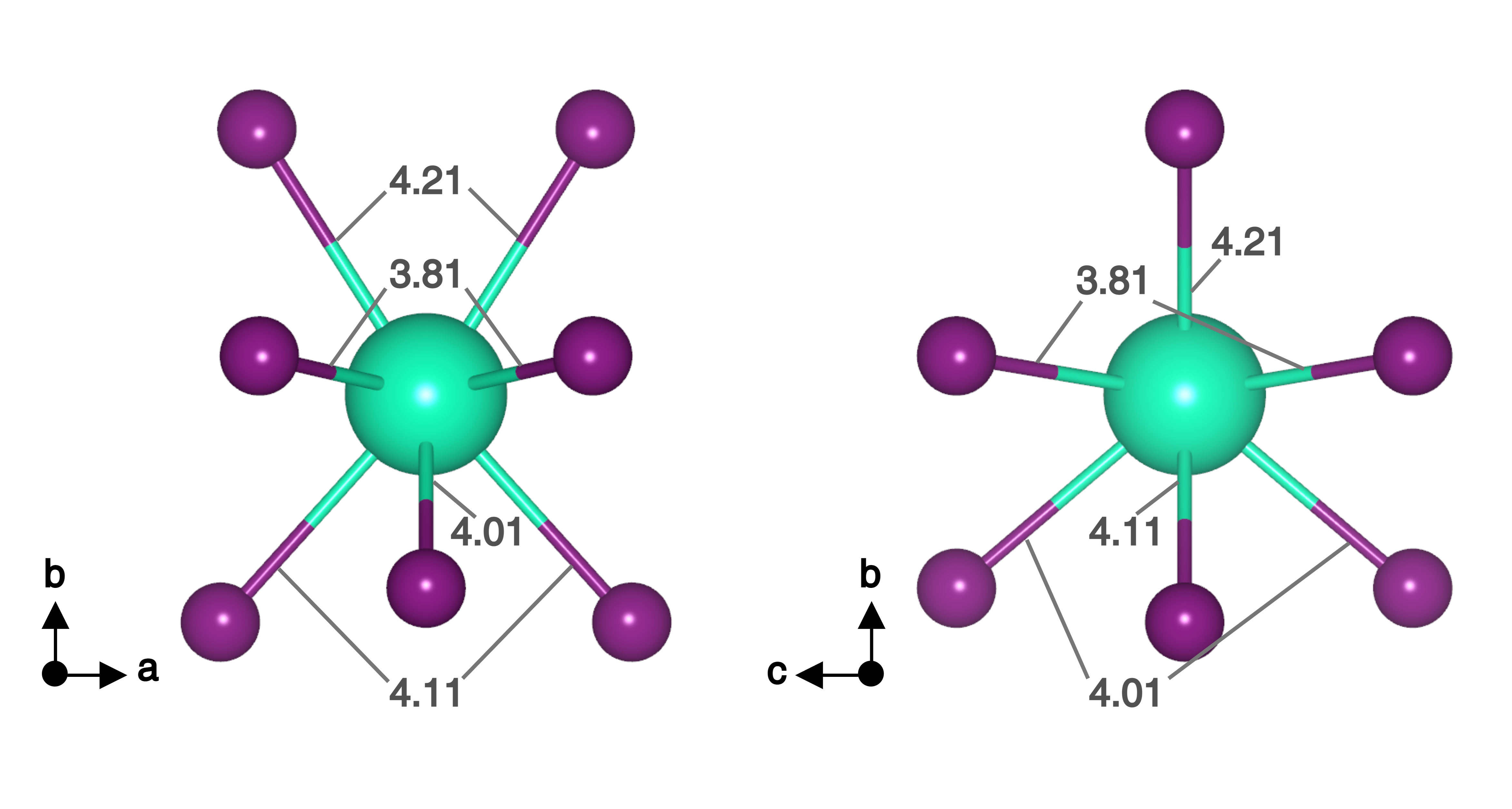}
  \caption{Cs--I interatomic distances in \ce{CsCu2I3}, $Amm2$ structure, projected to the $ab$- (left) and $bc$-plane (right) of the conventional unit cell.}
  \label{figure_S2}
\end{figure*}

%
\begin{figure*}[h]
  \includegraphics[width=1.0\textwidth]{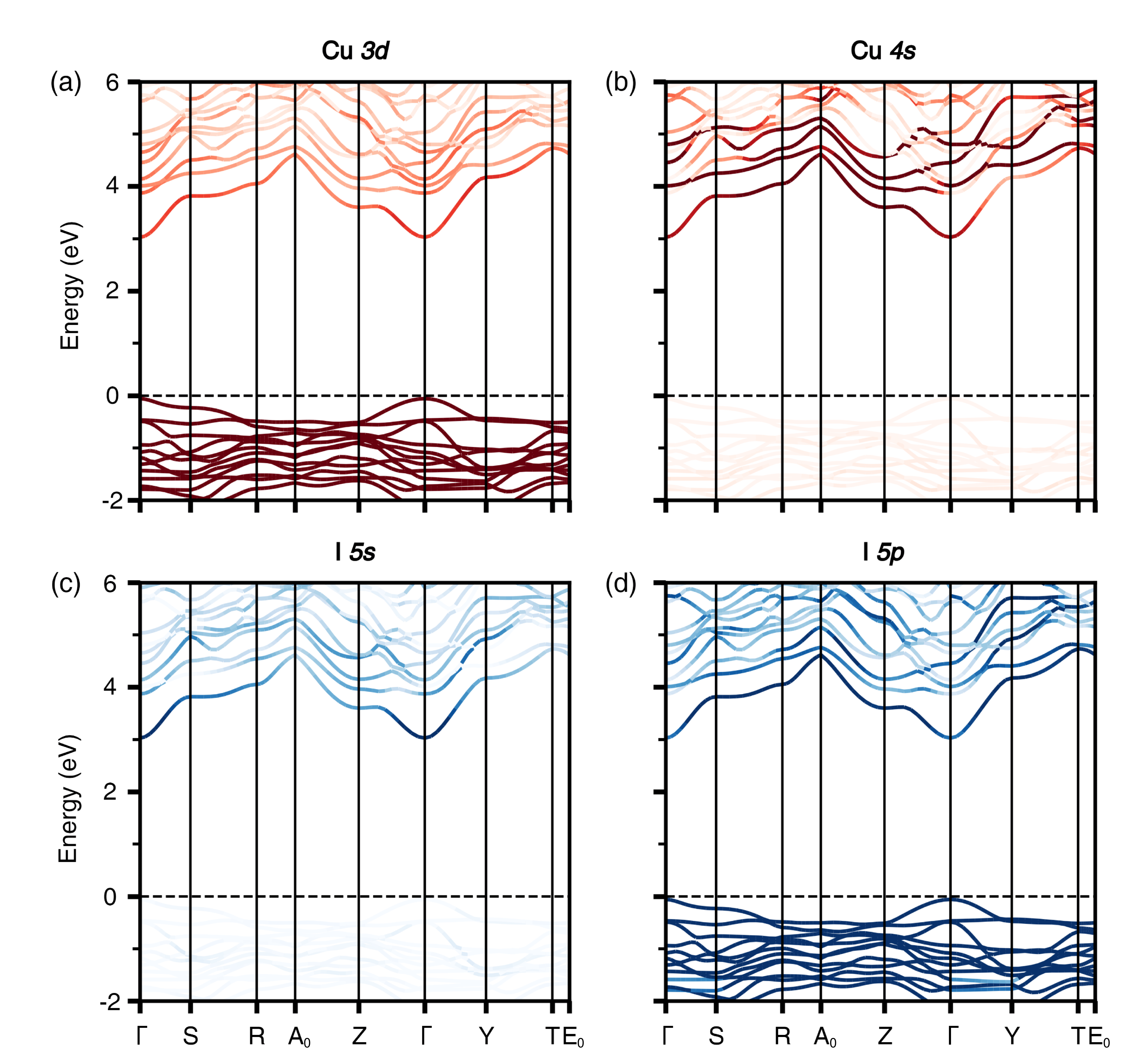}
  \caption{Electronic band structure of \ce{CsCu2I3}, $Amm2$ structure, divided by its individual orbital contribution; (a) Cu 3$d$ orbital; (b) Cu 4$s$ orbital; (c) I 5$s$ orbital; (d) I 5$p$ orbital. The intensity of the colors indicate the relative contribution to the structure.
}
  \label{figure_S3}
\end{figure*}

%
\begin{figure*}[h]
\includegraphics[width=1.0\textwidth]{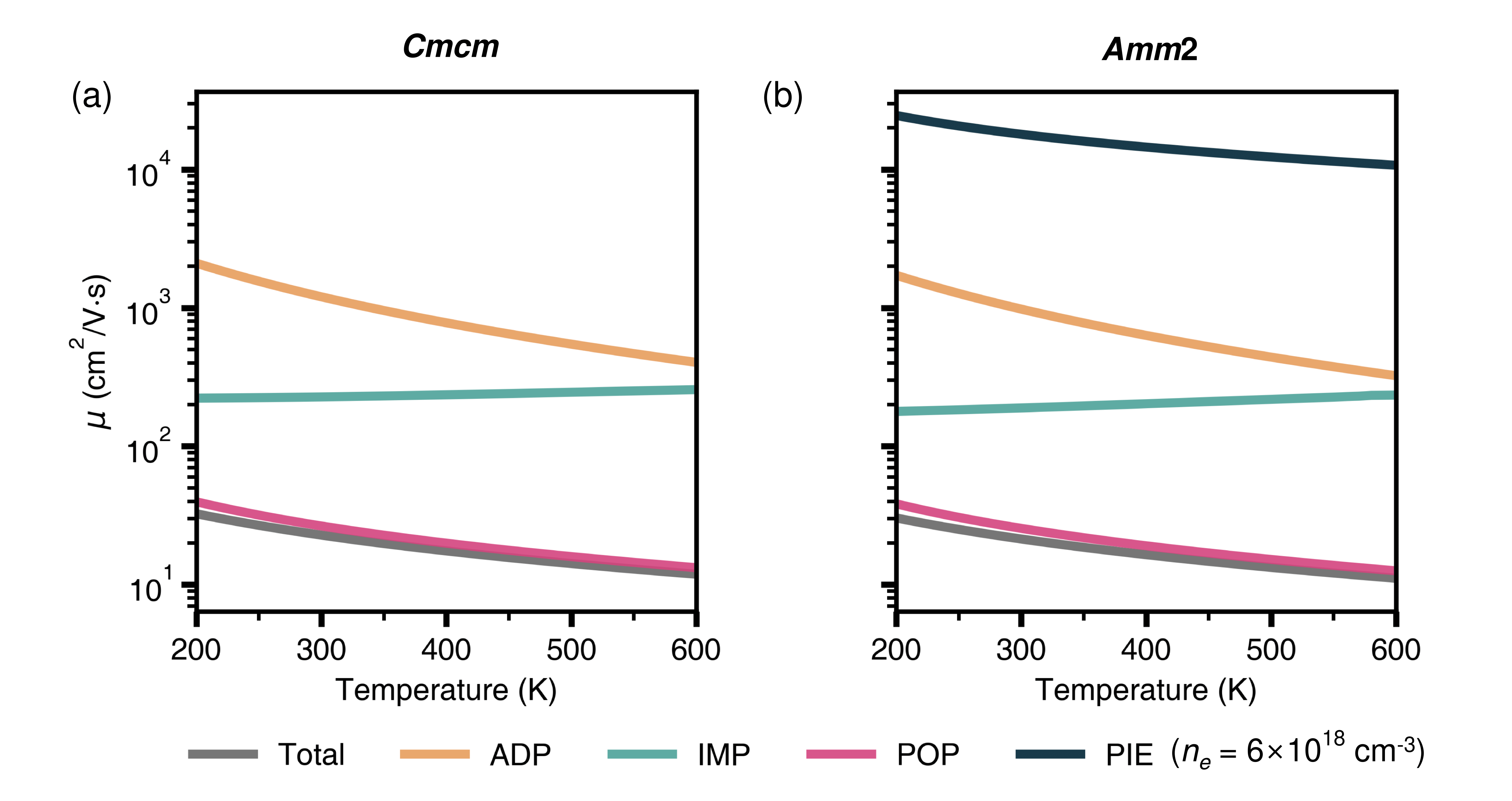}
\caption{Calculated electron mobility, $\mu$, as a function of temperature of \ce{CsCu2I3}, (a) $Cmcm$ and (b) $Amm2$ structure. Total mobility (gray) divided by its individual scattering mechanisms are also illustrated (acoustic deformation potential (orange, ADP), ionized impurity (teal, IMP), polar optical phonon (pink, POP), and piezoelectric scattering (black, PIE)). Electron concentration = $6\times10^{18}$\,cm$^{-3}$.}
  \label{figure_S4}
\end{figure*}

%
\begin{figure*}[h]
\includegraphics[width=1.0\textwidth]{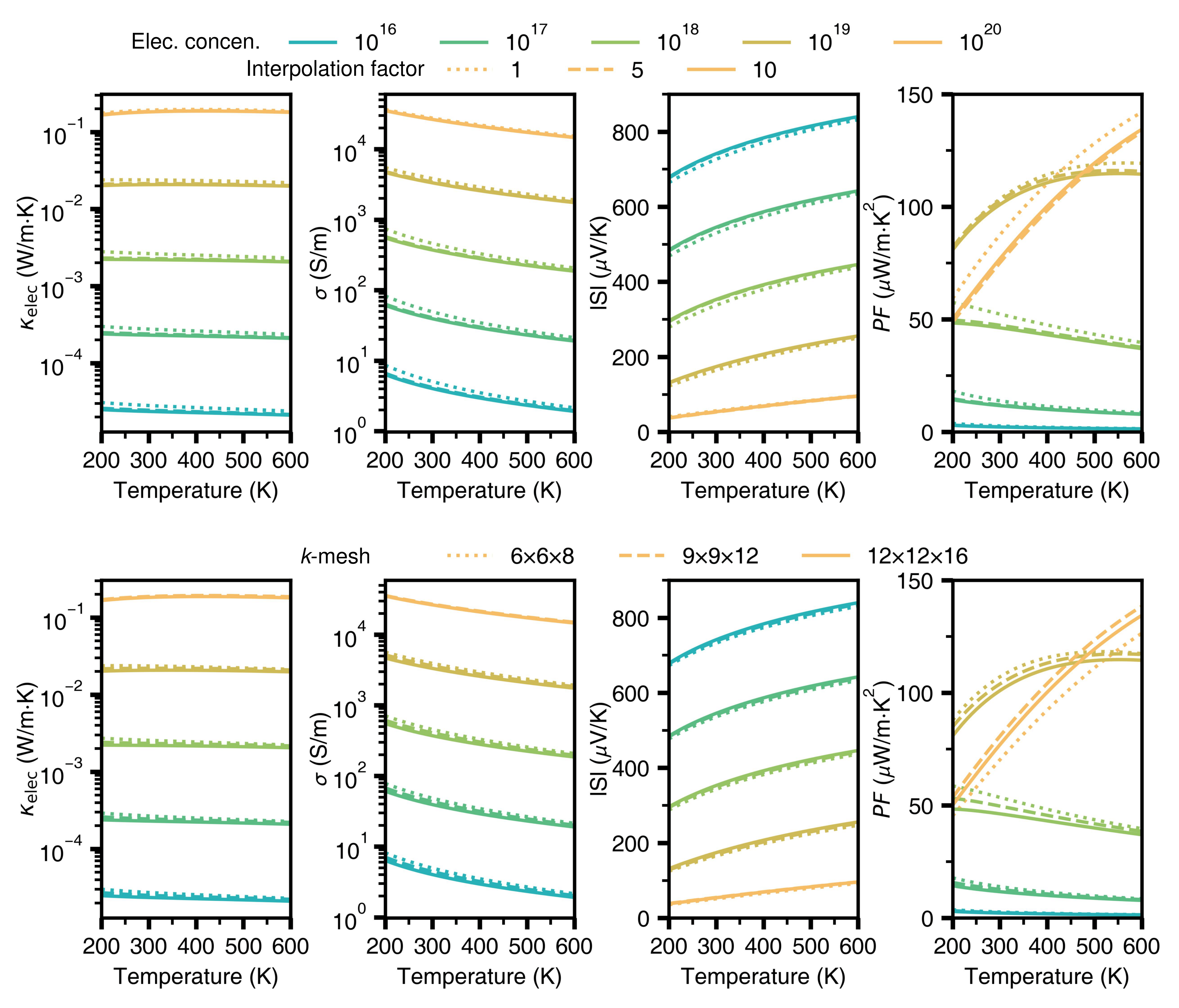}
\caption{Convergence test for calculated transport and thermoelectric properties of \ce{CsCu2I3}, $Amm2$ structure. Calculations over different interpolation factors with $k$-mesh fixed to $12\times12\times16$ (upper), and different $k$-meshes with interpolation factor fixed to 10 (lower).}
  \label{figure_S5}
\end{figure*}

%
\begin{figure*}[h]
\includegraphics[width=1.0\textwidth]{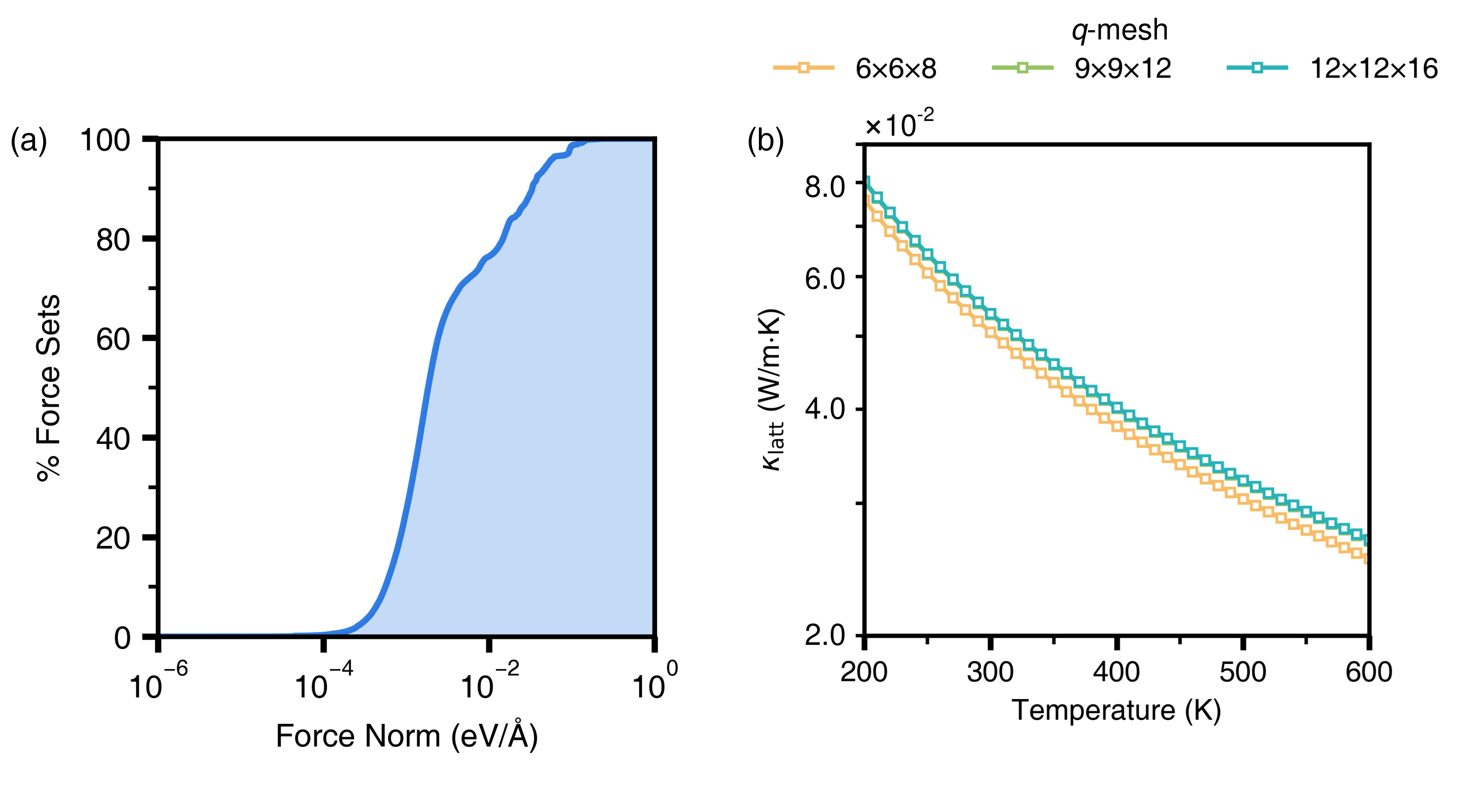}
\caption{Convergence test for calculated lattice thermal conductivity ($\kappa_{\rm latt}$) of \ce{CsCu2I3}, $Amm2$ structure. (a) Distribution of force norms for the force sets used for the calculation. All force norms are under $10^0$\,eV/{\AA}, proving its convergence. (b) Calculations over different $q$-mesh.}
  \label{figure_S6}
\end{figure*}
